\documentclass[10pt]{article}%
\usepackage{amsmath}
\usepackage{amssymb}
\usepackage{graphicx}
\usepackage{cite}
\usepackage{color}
\usepackage[labelfont=bf,labelsep=period,justification=raggedright]{caption}
\usepackage{amsfonts}%
\providecommand{\U}[1]{\protect\rule{.1in}{.1in}}
\makeatletter
\renewcommand{\@biblabel}[1]{\quad#1.}
\makeatother
\begin{document}

\date{}

\begin{flushleft}
{\Large \textbf{High-Fidelity Coding with Correlated Neurons}}

\bigskip

\bigskip

\textsc{Rava Azeredo da Silveira}$^{1,2,3,\ast}$, \textsc{Michael J. Berry
II}$^{4,\dagger}$ \newline

\bigskip

1\textit{\ Department of Physics, Ecole Normale Sup\'{e}rieure, 24 rue
Lhomond, 75005 Paris, France\newline}2\textit{\ Laboratoire de Physique
Statistique, Centre National de la Recherche Scientifique, Universit\'{e}
Pierre et Marie Curie, Universit\'{e} Denis Diderot, France}

3\textit{\ Princeton Neuroscience Institute, Princeton University, Princeton,
New Jersey 08544, U. S. A.\newline}4\textit{\ Department of Molecular Biology,
Princeton University, Princeton, New Jersey 08544, U. S. A. \newline}$\ast
$\textit{\ E-mail: rava@ens.fr \newline}$\dagger$\textit{\ E-mail:
berry@princeton.edu}

\bigskip
\end{flushleft}

\section*{Abstract}

Positive correlations in the activity of neurons are widely observed in the
brain. Previous studies have shown these correlations to be detrimental to the
fidelity of population codes or at best marginally favorable compared to
independent codes. Here, we show that positive correlations can enhance coding
performance by astronomical factors. Specifically, the probability of
discrimination error can be suppressed by many orders of magnitude. Likewise,
the number of stimuli encoded---the capacity---can be enhanced by similarly
large factors. These effects do not necessitate unrealistic correlation values
and can occur for populations with a few tens of neurons. We further show that
both effects benefit from heterogeneity commonly seen in population activity.
Error suppression and capacity enhancement rest upon a pattern of correlation.
In the limit of perfect coding, this pattern leads to a `lock-in' of response
probabilities that eliminates variability in the subspace relevant for
stimulus discrimination. We discuss the nature of this pattern and suggest
experimental tests to identify it.

\bigskip

\section*{Author Summary}

Traditionally, sensory neuroscience has focused on correlating inputs from the
physical world with the response of a single neuron. Two stimuli can be
distinguished solely from the response of one neuron if one stimulus elicits a
response and the other does not. But as soon as one departs from extremely
simple stimuli, single-cell coding becomes less effective because cells often
respond weakly and unreliably. High fidelity coding then relies upon
populations of cells, and correlation among those cells can greatly affect the
neural code. While previous theoretical studies have demonstrated a potential
coding advantage of correlation, they allowed only a marginal improvement in
coding power. Here, we present a scenario in which a pattern of correlation
among neurons in a population yields an improvement in coding performance by
several orders of magnitude. By \textquotedblleft
improvement\textquotedblright\ we mean that a neural population is much better
at both distinguishing stimuli and at encoding a large number of them. The
scenario we propose does not invoke unrealistic values of correlation. What is
more, it is even effective for small neural populations and in subtle cases in
which single-cell coding fails utterly. These results demonstrate a previously
unappreciated potential for correlated population coding.

\bigskip

\section*{Introduction}

Many of the classic studies relating behavior to the activity of neurons, such
as studies of single photon counting, have focused on behaviors that are near
the threshold of perception \cite{barlow_levick_1969, hecht_pirenne_1942,
klein_levi_1985, newsome_movshon_1989, watson_robson_1983}, where performance
is uncertain and can suffer a substantial error rate. One of the great
surprises of these studies is that in this limit, the variability of single
neurons often matches the variability in performance, such that single neurons
can account for the behavior \cite{Barlow_Yoon_1971, newsome_movshon_1989,
parker_hawken_1985}. However, most of our everyday visual experience involves
judgments made with great accuracy and certainty. As is illustrated by phrases
like \textquotedblleft seeing is believing\textquotedblright\ and
Shakespeare's \textquotedblleft ocular proof,\textquotedblright\ we often
dismiss any doubt about an aspect of the world once it is perceived visually.
In this `high-fidelity' limit, perception must cope with single neuron
variability by relying upon populations of neurons. Our visual system not only
yields perception with certainty, but it also allows us to make complex
judgments very rapidly---a fact that places additional constraints on the
population neural code \cite{kirchner_thorpe_2006, liu_kreiman_2009}.

In a neural population, correlations in the activity of neurons provide
additional variables with which information can be represented. While details
may vary from one neural circuit to another, a fairly common pattern of
correlation is observed across many brain regions, including the retina, LGN,
cerebral cortex, and cerebellum \cite{hatsopoulos_donoghue_1998,
mastronarde_1989, ozden_wang_2008, perkel_moore_1967, sasaki_llinas_1989,
shlens_chichilnisky_2008, usrey_reid_1999, vaadia_aertsen_1995}. Correlations
vary from pair to pair, with a positive mean and a standard deviation
comparable to the mean \cite{bair_newsome_2001, fiser_weliky_2004,
kohn_smith_2005, lee_georgopoulos_1998, zohary_newsome_1994} (but see Ref.
\cite{ecker_tolias_2010}).

How do these affect coding? This question has been investigated by a number of
authors \cite{johnson_1980, vogels_1990, oram_sengpiel_1998,
abbott_dayan_1999, panzeri_rolls_1999, sompolinsky_shamir_2001,
wilke_eurich_2002, romo_salinas_2003, golledge_young_2003, pola_panzeri_2003,
averbeck_lee_2003, shamir_sompolinsky_2004, shamir_sompolinsky_2006,
averbeck_lee_2006, averbeck_pouget_2006, josic_delarocha_2009}, who find that
in many cases positive correlations are detrimental to coding performance; in
some cases, however, positive correlations can enhance the coding performance
of a neural population. Using specific choices of neural response and
correlation properties, this effect was probed quantitatively in models of
pairs of neurons, small populations, or large populations. In all these cases,
the presence of correlation boosted coding performance to a relatively modest
degree: the mutual (Shannon) information or the Fisher information (depending
on the study) in the correlated population exceeded that in the equivalent
independent population by a factor of $O\left(  1\right)  $. For typical
choices of correlation values, the improvement was calculated to be
$\sim1\%-20\%$. These results can be translated into the units of capacity
used in this study and correspond to an improvement of a fraction of a percent
to a few percents (see Discussion below), which in turn correspond to a
negligible increase of the information encoded per neuron.

A limited number of experimental results indicate that information about
stimuli can be represented by the correlations themselves, but in the more
typical case single-cell responses vary with the stimuli while correlation
values are stimulus-independent. In this context, all the models which display
an improvement in coding in the presence of correlation rely on the same
mechanism \cite{johnson_1980, oram_sengpiel_1998, abbott_dayan_1999,
petersen_diamond_2001, sompolinsky_shamir_2001, wilke_eurich_2002,
romo_salinas_2003, averbeck_lee_2006, averbeck_pouget_2006}, namely, that
correlation relegates the variability of neural response into a
non-informative mode. To be more specific, because of variability each
stimulus is represented by a distribution of response patterns in the
population, and the overlap between neighboring distributions results in
coding ambiguity. While positive correlations broaden response distributions
overall, in a heterogeneous system the broadening can occur along a
non-informative direction while distributions compress along directions that
matter in terms of potential ambiguity (Fig. 1B). As a result, correlation can
suppress ambiguity and, hence, enhance coding fidelity.

Here, we focus upon the typical case of stimulus-independent correlation. We
exploit this same basic mechanism but we apply it to neural populations of any
size. Our central result is that, when the true population effect is taken
into account, the enhancement in coding performance can be astronomical. In a
correlated population, discrimination errors can be suppressed by factors of
$10^{20}$\ (or even greater) and the information per neuron can be boosted by
factors of $10$\ (or even greater), as compared with an equivalent independent
population. We obtain these results in simple models that assume experimental
values of correlations and population size. In fact, astronomical enhancement
of coding fidelity occurs in populations as small as $\sim10$\ neurons and in
situations in which independent coding breaks down entirely. We derive these
results numerically and analytically, and we discuss then in light of a
`lock-in' limit of the basic mechanism: unlike in pairs of neurons, in a
larger population correlation is able to compress response distributions into
an effectively lower-dimensional object. Furthermore, we demonstrate that
physiological heterogeneity, ubiquitous in neural systems, systematically
enhances the effect of correlation. Finally, we discuss the statistical
plausibility of the occurrence of correlated high-fidelity coding in actual
neural populations, and we point to a strategy for testing our predictions experimentally.

\section*{Results}

Our results amount to the answers to two complementary questions. Given a pair
sensory stimuli, how well can a population of correlated neurons discriminate
between them? Or, more precisely, what is the discrimination error rate?
Conversely, given a discrimination error rate, what is the capacity of a
correlated population?\ That is, how many stimuli can it encode with tolerable
error? In natural situations, discrimination errors are exceedingly rare and,
hence, neural populations are expected to achieve very low error rates. (See
Supplementary Experimental Procedures 1.1 for a layout of the assumptions used
in our approach and Supplementary Text 2.1 for a detailed argument and
quantitative estimates of low error rates.) All our discussion is set in this
low-error regime.

\subsection*{Positive correlations can suppress discrimination error rates by
orders of magnitude}

We consider two stimuli or sets of stimuli, which we henceforth refer to as
Target and Distracter. These can be two specific stimuli (\textit{e.g.}, a
black cat and a tabby cat), a specific stimulus and a stimulus category
(\textit{e.g.}, your cat and all other cats), or two stimulus categories
(\textit{e.g.}, all black cats and all tabby cats) and they have to be
discriminated by the response of a neural population in a short time window
during which each neuron fires $0$ or $1$ spike. Each neuron is bound to
respond more vigorously on average either to Target or to Distracter. Thus, it
is natural to divide the $N$-neuron population into two pools of neurons
(\textquotedblleft Pool 1\textquotedblright\ and \textquotedblleft Pool
2\textquotedblright), each more responsive to one of the two stimuli, as it
has been done customarily in studies on stimulus discrimination (see,
\textit{e.g.}, \cite{newsome_movshon_1989}). For the sake of simplicity, in
this 2-Pool model we allocate $N/2$ neurons to each pool (Fig. 1A). We denote
by $k_{1}$ and $k_{2}$ the number of active neurons in Pools 1 and 2
respectively. We start with a symmetric case: neurons in Pools 1 and 2 respond
with firing rates $p$ and $q$ respectively to the Target and, conversely, with
firing rates $q$ and $p$ respectively to the Distracter. Moreover,
correlations in the activity of pairs of neurons may take different values
within Pool 1 ($c_{11}$), within Pool 2 ($c_{22}$), and across pools ($c_{12}%
$). We denote by $C_{ij}$ the bare values of pairwise correlation and by
$c_{ij}$ the normalized pairwise correlations. Normalized values are often
quoted in the literature and present the advantage of being bounded by $-1$
and $1$. (See Experimental Procedures for mathematical definitions.) While we
shall present most of our quantitative results for symmetric choices of the
parameters, our qualitative conclusions hold in general.

If $p$ is larger than $q$, Pool 1 consists of the neurons `tuned' to Target
while Pool 2 consists of the neurons `tuned' to Distracter. A useful visual
representation of the probability distributions of responses to Target and
Distracter makes use of contour lines (Fig. 1B). In the case of independent
neurons (with $c_{11}=c_{22}=c_{12}=0$), the principal axes of the two
distributions are horizontal and vertical, and their contour lines are nearly
circular unless $p$ or $q$ take extreme values. As a result, the overlap
between the two distributions tends to be significant (Fig. 1B), with the
consequence of a non-negligible coding error rate. In such a situation,
positive correlations can improve coding by causing the distributions to
elongate along the diagonal and, conversely, to shrink along the line that
connects the two centers (Fig 1B).

To illustrate this generic mechanism, we have computed the error rate
numerically for specific choices of parameters of the firing rates and
correlations in the population. (See Supplementary Experimental Procedures 1.2
for a review of the maximum likelihood error and Supplementary Experimental
Procedures 1.3 for details on the numerics.) By way of comparison, in an
independent population with $N$ neurons the error rate drops exponentially as
a function of $N$ (Fig. 2A). While the error rates for independent and
correlated populations start out very similar for small population size, they
diverge dramatically as $N$ increases to~90 neurons (Fig. 2A). We can define a
factor of coding improvement due to correlations as the ratio of the two error
rates; this factor exceeds $10^{20}$ for large populations (Fig. 2B). We can
also explore the way in which the error rate changes as we vary the strength
of the pairwise correlations at fixed population size. Increasing the strength
of correlation across pools, $c_{12} $, sharply reduces the error rate, while
increasing the strength of correlation within pools, $c_{11}$ or $c_{22}$,
enhances the error rate (Figs. 2C and D).

The important point, here, is that improvements by orders of magnitude do not
result from growing the population to unrealistically large numbers of neurons
or boosting the correlations to limiting values. The massive suppression of
error rates occurs in populations of less than a hundred neurons and in the
presence of realistic correlations ranging from $c\approx$ 0.01 to 0.03. This
is because even in populations of relatively modest size, weak correlations
can significantly deform the shape of the probability distributions of
population responses (Fig. 2E).

In fact, the suppression of the coding error down to negligible values by
positive correlation does not even require populations with as many as
$N~\approx100$ neurons. Such suppression can be obtained in much smaller
populations, with a total number of neurons, $N$, between 8 and 20 and with
values of correlations below or not much higher than $c\approx0.3$ (Figs. 3A
and B). Such values of correlations are still well within the experimentally
measured range. We also explore another case which, naively, prohibits
low-error coding: that in which the firing rates in the two neuron pools
differ by very little; specifically, when $N\left(  p-q\right)  $ is of order
one. This condition implies that the overall activities in a given pool, in
response to Target and Distracter respectively, differ by one or a few spikes.
In this limiting case, coding with pools of independent neurons fails
entirely, with error rates of order one, since the absolute amplitude of
fluctuations exceeds unity. In a correlated population, we find, again, a
massive suppression of error rates by orders of magnitude, for realistic
values of correlation (Figs. 3C and D).

\subsection*{Analysis of Low-Error Coding}

In addition to our direct numerical investigations, we have performed analytic
calculations using a Gaussian approximation of the probability distribution
(see Supplementary Experimental Procedures 1.4 for derivations). The analytic
results agree very closely with the numeric results (Figs. 2 and 3, solid line
\textit{vs}. circles) and yield simple expressions for the dependence of the
error rate upon the parameters of our model, useful for a more precise
understanding of the effect of correlation.

The analytic expression of the error rate, $\varepsilon$, reads
\begin{equation}
\varepsilon=\frac{e^{-N\left(  p-q\right)  ^{2}/2\Delta}}{\sqrt{\pi N\left(
p-q\right)  ^{2}/\Delta}}. \label{error-rate-analytical}%
\end{equation}
The numerator in the argument behaves as expected for a population of
independent neurons: it yields an exponential decay of the error rate as a
function of $N$, with a sharpness that increases with the difference between
$p$ and $q$. But the denominator,%
\begin{align}
\Delta &  =p\left(  1-p\right)  \left\{  1-c_{11}+\frac{N}{2}\left[
c_{11}-\sqrt{\frac{q\left(  1-q\right)  }{p\left(  1-p\right)  }}%
c_{12}\right]  \right\} \nonumber\\
&  +q\left(  1-q\right)  \left\{  1-c_{22}+\frac{N}{2}\left[  c_{22}%
-\sqrt{\frac{p\left(  1-p\right)  }{q\left(  1-q\right)  }}c_{12}\right]
\right\}  , \label{denominator-delta}%
\end{align}
provides a strong modulation as a function of correlations (Figs. 2 and 3). In
the symmetric case with $p=1-q$ and $c_{11}=c_{22}$, this expression
simplifies to%
\begin{equation}
\Delta=p\left(  1-p\right)  \left[  2\left(  1-c_{11}\right)  +N\left(
c_{11}-c_{12}\right)  \right]  . \label{denominator-delta-special-case}%
\end{equation}
This quantity approaches zero when $N\delta c\sim\mathcal{O}\left(  1\right)
$, where $\delta c=c_{12}-\frac{N-1}{N}c_{11}$. Thus, in a population of tens
or hundreds of neurons, it is sufficient that $c_{12}$ exceed $c_{11}$ by less
than a few percent for the coding error to become vanishingly small.

From Eq. (\ref{error-rate-analytical}), it is apparent that the error rate
converges rapidly to zero with decreasing $\Delta$, and has an essential
singularity at $\Delta=0$. For any well-defined probability distribution,
$\Delta$ remains non-negative, but it can take arbitrarily small values. When
correlations are such that $\Delta$ is small enough, we are in a regime of
high-fidelity coding. The error vanishes for $\Delta\rightarrow0$; in this
limit, the probability distributions corresponding to Target and Distracter
are both parallel and infinitely thin. The value of $\Delta$ alone does not
specify the geometry of the probability distributions entirely; even with
$\Delta=0$, there remain free parameters, namely, the angles along which the
elongated distributions lie in the $(k_{1},k_{2})$ plane (denoted $\phi$ in
Fig. 1B). In Supplementary Experimental Procedures 1.6, we demonstrate that
these additional parameters need not be fine-tuned for high-fidelity coding.
In fact, angles can vary by as much as $\sim40^{\circ}$ while the error rate
remains below $10^{-12}$.

\subsection*{Neural Diversity Is Favorable to High-Fidelity Coding}

The simplest version of the 2-Pool model, discussed hitherto, assigns
homogeneous firing rate and correlation values within and across each of the
two neural sub-populations. Similar homogeneity assumptions are frequent in
modeling and data analysis: while response properties vary from neuron to
neuron in data, average values are often chosen to represent a population as a
whole and to evaluate coding performances. It is legitimate, however, to ask
to what extent error rates are shifted in a more realistic setting which
includes neural diversity and, in fact, whether high-fidelity coding survives
at all in the presence of neuron-to-neuron heterogeneity. We find that not
only does it survive but that, in fact, neural diversity further suppresses
the error rate.

We generalized the 2-Pool model of a correlated population to include
neuron-to-neuron diversity, by randomly and independently varying the firing
rate and correlation values according to a Gaussian distribution with standard
deviation $\sigma$, measured as a fraction of the original value. We then
computed the error rate in this generalized model and compared it to the
corresponding quantity in the homogeneous 2-Pool model. We found that every
single instantiation of neural diversity yielded an improvement in the coding
performance (Figs. 4A and B). More diverse neural populations with larger
values of $\sigma$ display stronger suppressions of the error rate (Fig 4C).
As $\sigma$ increases, the suppression factor grows both in mean and in
skewness, so that a significant fraction of the instantiations of
heterogeneity yields a large improvement of the coding performance over the
homogeneous case (Figs. 4A \textit{vs.} B).

The degree of error suppression depends, of course, on how much correlation
reduces the error relative to the matched independent population in the first
place. For the population shown here, neuron-to-neuron variations on a range
commonly seen in experiments lead to a suppression of the error rate by a
factor of $\sim5$ on average and a factor of $\sim50$ for some instantiations
of the heterogeneity (Fig. 4B). The coding benefit of heterogeneity appears to
be a rather general phenomenon \cite{wilke_eurich_2002,
shamir_sompolinsky_2006, osbourne_bialek_2008}.

\subsection*{The Mechanism for High-Fidelity Coding and the `Lock-In'
phenomenon}

The mechanism of dramatic error suppression from positive correlations may be
explained in a general manner that does not invoke a specific model or
approximation. A powerful description is given in terms of the `macroscopic'
variances and covariances of the spike count within and across the two pools:
we call $\chi_{11}^{2}$ the variance in the spike count, $k_{1}$, within Pool
1, $\chi_{22}^{2}$ the variance in the spike count, $k_{2}$, within Pool 2,
and $\chi_{12}^{2}$ the covariance of spike counts across the two pools. (See
Fig. 1B for a visual definition of these quantities, Experimental Procedures
for mathematical definitions, and Supplementary Experimental Procedures 1.5
for derivations of the results discussed below.)

The variances of the probability distribution of the neural response in the
plane $\left(  k_{1},k_{2}\right)  $ take the form%
\begin{equation}
\varkappa_{\pm}^{2}\equiv\frac{1}{2}\left(  \chi_{11}^{2}+\chi_{22}^{2}%
\pm\sqrt{\left(  \chi_{11}^{2}-\chi_{22}^{2}\right)  ^{2}+4\chi_{12}^{4}%
}\right)  . \label{min-max-variances}%
\end{equation}
The angles along which these variances are measured can also be computed
similarly (see Supplementary Experimental Procedures 1.5). In the case of
positive correlation, the angle along which the distribution elongates
(\textit{i.e.}, the angle long which $\varkappa_{+}$ extends, denoted $\phi$
in Fig. 1B) lies between $0^{\circ}$ and $90^{\circ}$. The small variance,
$\varkappa_{-}$, lies at right angle and governs error rate suppression. The
smaller $\varkappa_{-}$ and the more parallel the compressed distributions,
the smaller the error rates. The expressions for the variances (above) and the
angles (given in the Supplementary Experimental Procedures 1.5) are
general---they do not depend upon the shapes of the distributions or the
details of the correlation among neurons---and they give a sense of the extent
to which probability distributions of the population response are deformed by
correlations. In the specific 2-Pool models we treated above, positive
correlations induce massive suppressions of the coding error rate. We expect
similar high-fidelity coding whenever the tails of probability distributions
fall off sufficiently rapidly.

The limiting case of an infinitely thin distribution occurs when
\begin{equation}
\chi_{11}\chi_{22}=\chi_{12}^{2}; \label{macro-lockin-condition}%
\end{equation}
in this case,
\begin{equation}
\varkappa_{+}=\sqrt{\chi_{11}^{2}+\chi_{22}^{2}}%
\end{equation}
and
\begin{equation}
\varkappa_{-}=0.
\end{equation}
We refer to Eq. (\ref{macro-lockin-condition}) as the `lock-in' condition.
When the cross-pool covariance becomes this large, the width of the
probability distribution vanishes and the dimensionality of the response space
is effectively reduced by one. In the case of homogeneous pools of neurons, we
can reformulate this condition using `microscopic' correlations, as
\begin{equation}
\left[  1+\left(  \frac{N}{2}-1\right)  c_{11}\right]  \left[  1+\left(
\frac{N}{2}-1\right)  c_{22}\right]  =\frac{N^{2}}{4}c_{12}^{2}
\label{micro-lockin-condition}%
\end{equation}
(see Experimental Procedures and Supplementary Experimental Procedures 1.5).
If the lock-in condition in Eq. (\ref{macro-lockin-condition}) (alternatively,
Eq. (\ref{micro-lockin-condition})) is satisfied and $\chi_{11}$ and
$\chi_{22}$ (alternatively, $c_{11}$ and $c_{22}$) are chosen such as to yield
compressed distributions that are parallel, then error rates vanish. (See
Supplementary Text 2.4 on the nature of the locked-in state.)

As we have seen above, even if the cross-pool correlation approaches this
lock-in limit without achieving it, still the error rate can be suppressed
dramatically. Furthermore, the angles of the two distributions need not be
precisely equal. Thus, this amounts to a robust mechanism by which coding and
discrimination may be achieved with near-perfect reliability. It does not
require fine tuning of the parameters such the distribution widths and their
tilt angles; in particular, we need not limit ourselves to symmetric choices
of parameters, as we have done above for the sake of simplicity.

The general arguments presented here also indicate that the `$0$ or $1$ spike'
assumption is inessential and, in fact, that relaxing it may lead to even
stronger effects. If individual neurons can fire several spikes in a time
window of interest, the code can be combinatorial, but a simple spike count
code will do \textit{at least as well} as a more sophisticated combinatorial
one. If we stick to the spike count code, the general formulation remains
valid. In this situation, allowing many spikes per neurons corresponds
effectively to increasing the total number of neurons and, hence, can yield
stronger effects for comparable correlation values.

\subsection*{Correlated Populations Can Code for Large Sets of Stimuli with
High Fidelity}

In most natural situations, the task of organisms is not to tell two stimuli
apart but rather to identify an actual stimulus among a wealth of other,
possibly occurring stimuli. Visual decoding must be able to assign a given
response pattern to one of many probability distributions, with low error. In
other words, any pair of probability distributions of neural activity,
corresponding to two stimuli among a large set of stimuli, must have little
overlap. Thus, the problem of low-error coding of a large set of stimuli
amounts to fitting, within the space of neural activity, a large number of
probability distributions, while keeping them sufficiently well separated that
their overlap be small.

It is easy to see pictorially why the presence of correlation is favorable to
the solution of this problem. The state of the 2-Pool model is specified by
the number of active neurons in Pools 1 and 2, $k_{1}$ and $k_{2}$
respectively. If neurons are independent, probability distributions
(corresponding to different stimuli) have a near-circular shape with variances
along the horizontal and the vertical axes of order $k_{1}$ and $k_{2}$ (Fig.
5A). As a result, the only way to prevent tails from overlapping too much is
to separate the peaks of the distributions sufficiently. By contrast, since
correlated distributions are elongated, their centers can be placed near each
other while their tails overlap very little (Fig. 5B). Thus, many more
correlated distributions than independent distributions can be packed in a
given region in the space of neural responses (Figs. 5A and B).

We call $\Omega$ the maximum number of stimuli that a population of neurons
can code with an error rate less than $\varepsilon^{\ast}$ in the
discrimination of any stimulus pair. In the case of independent neurons (Fig.
5A), a simple calculation yields
\begin{equation}
\Omega_{\text{2-Pool}}^{\text{independent}}\lesssim\frac{2N}{\ln\left(  4/\pi
N\varepsilon^{\ast2}\right)  },
\end{equation}
where we have chosen the value of the error threshold to be small enough that
$\pi N\varepsilon^{\ast2}<4$ (see Supplementary Experimental Procedures 1.7
for derivations). In the correlated case (Fig. 5B), distributions are
elongated and, provided the correlations values are chosen appropriately,
error rates become vanishingly small even if the average firing rates of
nearby distributions differ by no more than a few, say $a$, spikes. We then
obtain
\begin{equation}
\Omega_{\text{2-Pool}}^{\text{correlated}}\approx\frac{N}{2a},
\end{equation}
since distribution centers can be arranged along a line that cuts through the
space of responses---a square with side $N/2$\ in the positive $\left(
k_{1},k_{2}\right)  $\ quadrant. (Note that more than one row of distributions
may be fitted into the response space of the neural populations if the
distributions are not too broad in their elongated direction, with a resulting
enhancement of $\Omega_{\text{2-Pool}}^{\text{correlated}}$. Figure 5B
illustrates a case in which three rows are accommodated. We do not include
these extra encoded stimuli in our calculations, thus remaining more
conservative in our estimate of coding capacity.) According to our earlier
results (Fig. 3D), even in moderately small populations the error rate becomes
exceedingly small for realistic choices of the correlation values when the
distribution centers are two spikes away from each other. Thus, we can choose
the value $a=2$\ to obtain an estimate of $\Omega_{\text{2-Pool}%
}^{\text{correlated}}$. Putting all this together, find that for low enough
$\varepsilon^{\ast}$\ correlated coding always wins over independent coding
(Fig. 5C) because $\Omega_{\text{2-Pool}}^{\text{independent}}$\ depends upon
$\varepsilon^{\ast}$\ much more strongly than $\Omega_{\text{2-Pool}%
}^{\text{correlated}}$\ does. Furthermore, in the limit of small error
thresholds, increasing the population size yields only a negligible
enhancement of capacity in the case of independent neurons, whereas in the
correlated case the number of faithfully encoded stimuli grows with population
size (Fig. 5D).

\subsection*{Positive Correlations in a Diverse Neural Population Can Enhance
Capacity by Orders of Magnitude}

Our arguments suggest that we ought to examine the behavior of the capacity of
heterogeneous neural populations because a greater degree of heterogeneity
amounts to higher dimensional versions of the situations depicted in Figs. 5A
and B, as we explain now. We define the $D$-Pool model: a heterogeneous
generalization of the 2-Pool model in which the neural population is divided
into $D$ sub-populations. As before, firing rates and correlations are
homogeneous within each pool and across pool pairs. For the sake of
simplicity, we consider symmetric pools with $N/D$ neurons each; we also
expect this arrangement to be optimal for coding. The state of the model is
completely defined by the number of active neurons in each pool.

In order to estimate $\Omega$, we have to examine how probability
distributions corresponding to different stimuli can be fitted within a
$D$-dimensional box enclosing $\left(  \frac{N}{D}\right)  ^{D}$ neural
states. And overlaps among distributions have to respect the prescribed error
rate threshold. In the case of independent neurons we have to fit in
$D$-dimensional near-circular objects, whereas in the case of correlated
neurons we have to fit in slender objects. It is intuitive that it is easier
to pack cucumbers in a box than to pack melons of a comparable volume, because
a greater amount of empty space is wasted in the case of spherical objects
such as melons, and indeed we find here that a greater number of correlated
distributions, as compared to independent distributions, can be packed in the
space of responses. The calculation gives%
\begin{equation}
\Omega_{D\text{-Pool}}^{\text{independent}}\lesssim\left[  \frac{4N}%
{D\ln\left(  2D/\pi N\varepsilon^{\ast2}\right)  }\right]  ^{D/2}%
\end{equation}
(Fig. 6A, see Supplementary Experimental Procedures 1.7 for derivations).
Notice that the number of possible stimuli encoded by the independent
population increases for greater heterogeneity (larger $D$).

In the case of correlated neurons, distributions may be compressed along one,
two, ..., or $D-1$\ directions. In the latter case, indeed the most favorable
scenario, we have to pack near-one-dimensional objects. As before in the case
of $2$-Pools, we can assume that neighboring distributions centers are
separated by $a$\ spikes, and we obtain%

\begin{equation}
\Omega_{D\text{-Pool}}^{\text{correlated}}\approx\left(  \frac{N}{Da}\right)
^{D-1}.
\end{equation}
This simple result follows from the observation that distribution centers can
be arranged on a hyperplane that cuts through the hypercube of the space of
responses (see Supplementary Experimental Procedures 1.8 for a more detailed
discussion an a slightly more careful bound). From these expressions we can
conclude that the enhancement in capacity due to correlation is significant,
and that the enhancement increases with the degree of heterogeneity (Fig. 6B).

The number of stimuli encoded with tolerable error rate, $\Omega$, scales
differently with model parameters in the independent and correlated cases. In
order to focus on this scaling behavior, we define the `capacity per neuron',
$\mathcal{C}$, by analogy to the information conveyed by each neuron in a
population of perfectly deterministic neurons. In the latter case, the
population has access to $2^{N}$ response patterns that can code for stimuli
with perfect reliability. Each neuron conveys $\log_{2}\left(  2^{N}\right)
/N=1$ bit of information. Consequently, we define the capacity per neuron as%
\begin{equation}
\mathcal{C}\equiv\frac{\log_{2}\left(  \Omega\right)  }{N}.
\end{equation}
It is a measure of the mutual (Shannon) information per neuron in the
population in the limit of very small $\varepsilon^{\ast}$.

To explore the scaling behavior of correlated \textit{versus} independent
populations, it is reasonable to ask what degree of heterogeneity, as measured
by $D$, maximizes $\mathcal{C}$ for each value of $N$. Equivalently, we can
ask what pool size, $n\equiv N/D$, maximizes $\mathcal{C}$ (Fig. 6C, see
Supplementary Experimental Procedures 1.7 and 1.8). In the correlated case,
the optimal capacity obtains when heterogeneity is strong, in fact so strong
that the number of neurons per pool,~$n$, is as small as $5 $\ to
$10$\ neurons for the choice $a\approx1-2$. From the optimal pool size, we
find that the optimal value of the capacity per neuron is given by%
\begin{equation}
\mathcal{C}_{D\text{-Pool, optimal}}^{\text{independent}}\lesssim\left[
e\ln\left(  2\right)  \ln\left(  \frac{4}{\pi e\varepsilon^{\ast2}}\right)
\right]  ^{-1}%
\end{equation}
and%
\begin{equation}
\mathcal{C}_{D\text{-Pool, optimal}}^{\text{correlated}}\gtrsim0.28\text{
}\left(  0.14\right)  \text{\qquad for\qquad}a\approx1\text{ }\left(
2\right)
\end{equation}
in the independent and correlated cases respectively (see Supplementary
Experimental Procedures 1.7 and 1.8 for derivations). The independent capacity
becomes very small at low-error thresholds, while the correlated capacity
remains fixed and in fact of the same order as the capacity of a perfectly
reliable neuron (Fig. 6D). Thus, in the limit of low error, the capacity and
hence information encoded per neuron exceeds the corresponding quantity in an
independent population by more than a factor of $10$. By comparison, one often
finds analogous effects measured in a few percent in other studies.

We have put forth the following picture. For a neural population to code for a
large set of inputs reliably, it breaks up into small pools with about ten
neurons, with correlation across pools stronger than correlation within pools.
These pools are small enough that their number is large, and consequently the
response space is high-dimensional. But, at the same time, the pools are large
enough that realistic correlations lock them in and yield effectively
lower-dimensional response distributions.

\subsection*{Experimental Prediction and Plausibility of Correlated
High-Fidelity Coding}

If neural populations rely upon correlation to achieve high-fidelity coding,
we expect that patterns of correlations resembling those postulated in our
model can be found in data. Namely, our hypothesis predicts that subsets of
similarly tuned pools of neurons will exhibit weaker within-pool correlations
than cross-pool correlations. In order to check this prediction, the response
of a neural population to a pair of stimuli or a pair of stimulus classes has
to be recorded (Fig. 7A). This population is divided into a group of cells
that fire more strongly to the first stimulus and the rest that fire more
strongly to the second stimulus (Fig. 7B). Note that this step is always
possible and that all cells can be thus assigned.

One then searches for subsets of the population that have stronger correlation
across the groups than within (Fig. 7C). For recordings with several tens of
cells, there is a very large number of possible subsets, so an exhaustive
search may not be feasible. Instead, there exist a number of faster search
strategies. For instance, one can score each cell according to the sum of its
pairwise correlation to all cells in the other group minus the sum to all
cells within its stimulus-tuned group. This yields a rank ordering of cells,
which can be used for selecting favorable subsets. In addition, searches can
be made iteratively, starting with $M$ cells and finding the best next cell to
add to the subset. Once a subset is identified, a quick assessment of the role
of correlation can be made using average firing rates and correlations to
calculate the error rate in the Gaussian approximation (Eq.
(\ref{error-rate-analytical})). As seen in Figs. 2 and 3, this approximation
is highly accurate. Then, for the most favorable subsets, a maximum entropy
calculation can be carried out to estimate the discrimination error taking
into account the true experimentally observed heterogeneity. As indicated by
Fig. 4, the homogeneous approximation is not only quite close to the real
error rate, but it also serves as an upper bound on the error. In this manner,
subsets of neurons with correlation patterns favorable to lock-in can be
identified in neurophysiological recordings.

A detailed analysis of neurophysiological data must await a subsequent study.
Here, we mention several observations which are consistent with our
experimental prediction. Patterns of correlations with stronger cross-pool
values may at first seem unlikely; this intuition comes mainly from our
knowledge of the primary visual cortex and area MT, in which neurons with
similar orientation tuning or directional preference are more strongly
correlated, on average. But recent results in the literature hint to the fact
that inverse patterns of correlation, with stronger cross-pool values, may
well be present in the brain and favorable to coding. Romo and colleagues have
reported precisely this phenomenon in S2 cortex: positive correlation among
pairs of neurons with opposite frequency-tuning curves
\cite{romo_salinas_2003}. This pattern of correlation resulted in an
improvement in the threshold for discrimination between different frequencies
of tactile stimulation. Maynard and Donoghue similarly found that a model that
incorporated correlation reduced discrimination errors, as compared to an
independent model, for groups of up to 16 cells in M1 during a reaching task
\cite{maynard_donoghue_1999}. Here, correlations elongated the response
distributions precisely in the manner depicted in Fig. 2B. Interestingly,
Cohen and Newsome observed that MT neurons with widely different direction
preferences displayed stronger positive noise correlation when the
discrimination task was designed in such a way that, effectively, they
belonged to different stimulus-tuned pools \cite{cohen_newsome_2008}. In
another study cortical study, Poort and Roelfsema demonstrated that noise
correlation can improve coding between V1 cells with different tuning,
partially canceling its negative effect on cells with similar tuning
\cite{poort_roelfsema_2009}. Finally, Gutnisky and Dragoi
\cite{gutnisky_dragoi_2008} observed that after rapid (400 ms) adaptation to a
static grating, pairwise correlation coefficients among neurons with similar
tuning decreased more than for neurons with different tuning preferences --- a
trend in adaptation which agrees with the proposed favorable pattern of correlation.

Because favorable patterns of correlation can dramatically reduce the coding
error even when they involve only a small number of neurons, the brain may
take advantage of the heterogeneity of pairwise correlations to read out from
small sub-populations which exhibit the proposed favorable pattern of
correlation. In order to evaluate the plausibility of this scenario, we ask:
If pairwise correlations are randomly distributed, with the experimental mean
and standard deviation, then how likely is it to find a favorable pattern of
correlation in a 2-Pool system with $M$ neurons in each pool within a local
population? By local population we mean, for example, a cortical column with
$N_{0}\approx10^{4}-10^{5}$ neurons. We find that a highly favorable pattern
of correlation is present with significant probability provided $N_{0}\geq
N_{0}^{\text{critical}}$, where the number $N_{0}^{\text{critical}}$ can be
estimated in terms of cortical parameters (see Supplementary Experimental
Procedures 1.9). Favorably correlated patterns with 8 neurons ($M=4$) occur
randomly with significant probability in local populations of no more than a
few hundred neurons, and their analog with 16 neurons ($M=8$) occur randomly
with significant probability in local populations of no more than a few
thousand neurons (see Supplementary Fig. S2A).

In a large population with $N_{0}\gg N_{0}^{\text{critical}}$ neurons, a great
number of 2-Pool systems that are close to lock-in are bound to occur
statistically (see Supplementary Experimental Procedures 1.9). Specifically,
we find that in an overall local population of 1000 neurons, which might be
thought of as the lower limit on the size of a cortical column, the number of
favorable patterns present ranges from a few tens (for patterns with $M=6$) to
several millions (for patterns with $M=4$) (see Supplementary Fig. S2B). Thus,
while they may be hard to identify experimentally, small locked-in systems may
populate cortical columns in a statistically significant manner.

\section*{Discussion}

We have shown that a class of patterns of positive correlation can suppress
\textit{coding errors} in a two-alternative discrimination task (Figs. 2A and
B). The idea that correlations among neurons may be favorable to coding was
noted earlier. What is new, here, is the demonstration of the extreme degree
of the enhancement in coding fidelity --- several orders of magnitude rather
than a few tens of a percent. Furthermore, this generic result does not
require unrealistic values of correlation or population size: it can operate
at the moderate values of correlations recorded experimentally (Figs. 2C and
D) and in populations with as few as $\sim10$ neurons (Figs. 3A and B). In
fact, massive error suppression may occur even when average activities in a
neural pool in response to different stimuli differ by one or a few spikes
(Figs. 3C and D)---a limiting, but realistic, situation in which coding with
independent neurons fails completely. The extreme nature of this effect makes
it more likely that the brain might deploy resources to create these favorable
correlation patterns.

We have also shown that correlations can boost dramatically the
\textit{capacity} of a neural population, \textit{i.e.}, the number of stimuli
that can be discriminated with low error (Figs. 5 and 6). For independent
neurons, the mean firing rates of the population in response to different
stimuli must differ by a substantial amount to allow low error, because the
firing variability about the mean is not harnessed by correlation. By
contrast, in the presence of correlation, neural response distributions can
deform into slender objects, effectively lower-dimensional objects, which can
be fitted much more efficiently within the population's response space (Fig.
5B). At lock-in, response distributions become strictly lower-dimensional
(one-dimensional in the extreme case).

Finally, we have demonstrated that diversity in neuron-to-neuron response, and
more generally heterogeneity of the population response, further enhances the
effect of correlation (Fig. 4 and Figs. 6A and B). Indeed, the advantageous
role of heterogeneity seems to be a rather general feature of population
coding, and it has been illustrated within various approaches
\cite{wilke_eurich_2002, shamir_sompolinsky_2006, osbourne_bialek_2008}. We
refer to the phenomenon in which neural correlation suppresses the
discrimination errors to negligible values and dramatically boosts the
capacity of a population as \textit{high-fidelity coding}. In passing, we note
that high-fidelity coding does not, in principle, require equal-time
correlation: the same mechanism can be at play when the correlations that
matter involve different time bins, such as in `spike-latency codes'
\cite{gollisch_meister_2008}.

\subsection*{Relation with Earlier Work on Coding with Correlation}

A number of theoretical studies have explored the role of correlation in
neural coding, with the use of different neuron models and information
theoretic measures \cite{johnson_1980, vogels_1990, oram_sengpiel_1998,
abbott_dayan_1999, panzeri_rolls_1999, sompolinsky_shamir_2001,
wilke_eurich_2002, romo_salinas_2003, golledge_young_2003, pola_panzeri_2003,
averbeck_lee_2003, shamir_sompolinsky_2004, shamir_sompolinsky_2006,
averbeck_lee_2006, averbeck_pouget_2006, josic_delarocha_2009}. If response
properties are homogeneous among neurons, positive correlation is detrimental
to coding: it tends to induce neurons to behave alike, and thereby suppresses
the advantage of coding with a population rather than with a single cell (see
Supplementary Text 2.2 for detailed arguments). By contrast, if response
properties vary among neurons, positive correlation can be either unfavorable
or favorable \cite{abbott_dayan_1999, panzeri_rolls_1999,
sompolinsky_shamir_2001, wilke_eurich_2002, pola_panzeri_2003,
averbeck_pouget_2006, shamir_sompolinsky_2006}. Put more generally, when the
scale of correlation is comparable to that of the informative mode in the
system (dictated, \textit{e.g.}, by the response tuning curve), then
correlation enhances the confounding effect of noise (see Supplementary Text
2.3 for a simple illustration of this mechanism). But when the scale and
structure of correlation is very different --- as in the case of uniform
positive correlations, in the case of negative correlations
(anti-correlations), or in models with heterogeneity --- correlation can
relegate noise to a non-informative mode \cite{abbott_dayan_1999,
sompolinsky_shamir_2001, wilke_eurich_2002}. (We recall that we are focussing
exclusively upon stimulus-\textit{independent} correlations. When correlations
depend upon the stimulus, they are evidently always favorable, as they
represent an additional variable that can play a role analogous to the mean
response \cite{panzeri_rolls_1999, wilke_eurich_2002, pola_panzeri_2003,
shamir_sompolinsky_2004, josic_delarocha_2009}. Experiments indicate the
presence of both stimulus-independent and stimulus-dependent correlations.)

In the case of stimulus-independent, positive correlation, earlier studies
have formulated a mechanism by which correlation can relegate noise to
non-informative models and, hence, enhance coding fidelity \cite{johnson_1980,
oram_sengpiel_1998, abbott_dayan_1999, petersen_diamond_2001,
sompolinsky_shamir_2001, wilke_eurich_2002, romo_salinas_2003,
averbeck_lee_2006, averbeck_pouget_2006}. Namely, that negative signal
correlations (anti-correlations) should be supplemented with positive noise
correlations. To be explicit, this means that when neurons respond
differentially to different stimuli, on average, then the variability about
this average response should be correlated positively; this mechanism is
illustrated in Fig. 1B and sets the stating point of our study. Conversely,
negative correlations (anti-correlations) are favorable in the case of
positive signal correlation. These statements have been established following
different routes in the literature. They can be read off in full generality,
that is, without invoking any particular neuron model or form of the neural
response, from the expression of the mutual (Shannon) information
\cite{panzeri_rolls_1999, golledge_young_2003, pola_panzeri_2003,
averbeck_pouget_2006}. This is done by rewriting the mutual information in a
form that displays contributions from firing rates, correlations, and the
interplay of firing rate and correlation patterns. Approaches using the mutual
information have the merit of elegance and generality. However, for
quantitative estimates they require the implementation of specific response
models; furthermore, they are difficult to apply to large populations of
neurons because of sampling limitations and mathematical difficulties.

Similar results can be derived from the form of the Fisher information
\cite{abbott_dayan_1999, sompolinsky_shamir_2001, wilke_eurich_2002,
averbeck_pouget_2006}, often used to establish bounds on the estimation
variability in the case of continuous stimuli. Most studies consider neurons
with broad tuning properties and find that positive correlations are
unfavorable if they decay on the scale of the tuning curve. Positive
correlations where observed to be favorable in cases in which they are uniform
among all neurons or have a non-monotonic profile according to which similarly
tuned neurons are less correlated than neurons that differ greatly in their
tuning. In all cases, however, positive correlation enhanced the coding
fidelity by modest amounts. In the next section, we discuss these quantitative
aspects in greater detail, as well as their correspondence with our
formulation and results.

In models of broadly tuned neurons with uniform pairwise correlation over the
entire population, coding becomes increasingly reliable as the quantity
$c$\ tends to $1$. For example, the Fisher information is boosted by a factor
$1/\left(  1-c\right)  $\ as compared to the case of independent neurons
\cite{abbott_dayan_1999}. Thus, strong correlation-induced improvement in
coding performance occurs only in the unrealistic limit of $c$\ close to $1$.
The situation is different in our simple models. There, high-fidelity coding
requires that the modified quantity $N\delta c$\ approach $1$, where $\delta
c$\ is a weighted difference of cross-pool correlation values and within-pool
values, be small (see, e.g., Eqs. (\ref{denominator-delta}) and
(\ref{denominator-delta-special-case})). The presence of similarly tuned pools
of neurons, within the population, amplifies the effect of weak pairwise
correlation to produce profound changes in the activity patterns of the neural
population. Since correlation values are in the range $c\approx1\%-30\%$,
values of $N$\ as modest as a few tens or a few hundreds are sufficient to
bring the quantity of interest, $N\delta c$, extremely close to $1$.
(Similarly, Sompolinsky et al. \cite{sompolinsky_shamir_2001} showed that
coding can be enhanced by a large factor in the presence of anti-correlations
as weak as $c=-0.005$ (as quoted, also, in Ref. \cite{averbeck_pouget_2006}).
This occurs for \ populations with $\sim500$\ neurons and it is yet another
illustration of the significant effect that can take place when $N\delta c\sim
O\left(  1\right)  $. In the present work, we have shown that similarly large
effects can occur due to the experimentally more typical positive
correlations, and in the context of much smaller neural population with no
more than a few tens of neurons.)

We remark in passing that there are other mechanisms by which confounding
noise can be relegated to non-informative dimensions. In the context of
broadly-tuned neurons and long-range correlation---the usual setup of studies
which make use of Fisher information---the presence of neuron-to-neuron
variability (e.g., in the firing rates) can do the trick
\cite{wilke_eurich_2002, shamir_sompolinsky_2006}. In the absence of
variability, positive correlation suppresses the coding performance as
compared with an independent population. Neuron-to-neuron variability
introduces a new dimension, namely, modulations much finer-grained than the
scale of tuning and correlation, in which information is stored. Then, in a
correlated population one retrieves, roughly, the coding performance of an
independent population. This mechanism cannot, to our knowledge, generate
substantial improvement in coding performance over that of an independent population.

\subsection*{Quantitative Aspects in Earlier and the Present Work --- Error
Rate and Capacity versus Shannon Information and Fisher Information}

As mentioned in the introduction and in the previous section, earlier
investigations which exhibit an improvement of the coding performance due to
positive correlation find that the latter is rather limited quantitatively.
Specifically, the Shannon information or the Fisher information (depending on
the study) in the correlated population exceed that in the equivalent
independent population by less than a factor of $O\left(  1\right)  $. As
stated above, the Fisher information can be boosted by a factor $1/\left(
1-c\right)  $\ as compared to its counterpart for a population of independent
neurons; for typical choices of correlation values, this yields an improvement
of $\sim1\%-20\%$. By contrast, in the present study we claim that positive
correlation can enhance coding fidelity by astronomical factors, and that this
effect exists even in small populations of neurons. But how are we to compared
our results to earlier results, since former are expressed in terms of error
rate and capacity while the latter are expressed in terms of information measures?

In the case of an unbiased estimator, the Fisher information, $I_{F}$, bounds
from below the discrimination error, $\rho$, of a continuously variable
stimulus: $\rho\geq1/\sqrt{I_{F}}$\ \cite{cover_thomas_1991}. Thus, if the
stimulus spans a space of size $L$\ then the number of stimuli that can be
distinguished reliably is calculated as%
\begin{equation}
\Omega\approx\frac{L}{\rho}\lesssim L\sqrt{I_{F}},
\end{equation}
so that the capacity per neuron scales with the Fisher information as
$C=\log_{2}\left(  \Omega\right)  /N\lesssim\log_{2}\left(  L\sqrt{I_{F}%
}\right)  /N$. (A rigorous version of this result was derived for a population
of independent neurons in Refs. \cite{brunel_nadal_1998,kang_sompolinsky_2001}%
.) If correlation enhances the Fisher information by a factor $\Delta I/I$,
$I_{F}^{\text{correlated}}=I_{F}^{\text{independent}}\left(  1+\Delta
I/I\right)  $, then the number of distinguishable stimuli is correspondingly
enhanced according to $\Omega^{\text{correlated}}\approx L\sqrt{I_{F}%
^{\text{correlated}}}=\Omega^{\text{independent}}\sqrt{1+\Delta I/I}$. Thus,
we have%
\begin{equation}
\frac{\Omega^{\text{correlated}}}{\Omega^{\text{independent}}}\approx
\sqrt{1+\frac{\Delta I}{I}},
\end{equation}
and%
\begin{equation}
\frac{\mathcal{C}^{\text{correlated}}}{\mathcal{C}^{\text{independent}}%
}\approx1+\frac{1}{N}\frac{\log_{2}\left(  1+\Delta I/I\right)  }%
{2\mathcal{C}^{\text{independent}}}%
\end{equation}
or%
\begin{equation}
\mathcal{C}^{\text{correlated}}-\mathcal{C}^{\text{independent}}\approx
\frac{1}{2N}\log_{2}\left(  1+\frac{\Delta I}{I}\right)  .
\end{equation}
We can now relate the earlier results in terms of Fisher information to our
results in terms of capacity through these formul\ae .

An enhancement of the Fisher information given by $\Delta I/I\sim O\left(
1\right)  $\ or, to be more specific, $\Delta I/I\approx0.01-0.2$\ as
suggested by earlier theoretical studies, amounts to a small increase of the
number of distinguishable stimuli by a factor $1.005-1.1$. Similarly, the
difference between correlated and independent capacity per neuron decays
inversely proportionately with $N$; in a large population, the improvement
becomes negligible. By contrast, we found that the ratio $\Omega
^{\text{correlated}}/\Omega^{\text{independent}}$\ can attain large values
($\approx3-10^{5}$, Fig. 6B) and that the difference between the correlated
capacity per neuron, $C^{\text{correlated}}$, and the independent capacity per
neuron, $C^{\text{independent}}$, can be significant (Fig. 6D). In brief,
earlier studies have demonstrated that, in spite of positive correlations,
coding can be as efficient as in an independent population or even slightly
better. Here, we show that, provided true population effects are taken into
account, positive correlation can have a profound quantitative effect in that
they can modulate the way coding measures scale with the number of neurons in
the population and, as a result, yield a massive enhancement in coding fidelity.

To conclude the comparison among information measures, we note that, for
continuous stimuli, Fisher Information is a natural performance metric. In
this case, stimulus entropy always exceeds that of the population response,
and the estimation variability decreases with population size, so that one is
interested in quantifying the precision of estimation in the large-$N$\ limit.
By contrast, here we treat the case of a discrete stimulus, where the entropy
is small and discrimination can be achieved with great reliability. This
regime is clearly relevant to tasks like decision-making, language, and
abstract thought: each categorization error imposes a cost on the organism,
making it relevant to characterize coding performance using the error rate
rather than the mutual information. Much of computational neuroscience work
devoted to networks of neuron has focused upon large-$N$\ situations. The
regime at hand here is somewhat new in character: the largest number is not
$N$, the population size, but rather $1/\varepsilon$, the inverse
discrimination error. In fact, a number of neurons as small as $N\sim10$\ can
achieve inverse error rate, $1/\varepsilon$, several orders of magnitude
larger. Given the breath and accuracy of cerebral function, and the brain's
limited size, we expect this regime to be relevant to diverse instances of
neural processing.

\subsection*{Other Strategies for Low-Error Coding}

As explained in Supplementary Text 1.1, infinitesimal error is not a luxury,
but a necessity in rapid coding if one wishes to avoid relatively frequent
false alarms. We have shown here how correlations can enable population codes
to perform with negligible error rates. However, other possible strategies for
reducing false alarm errors exist: temporal integration and prior expectation.
Both strategies effectively involve raising the detection threshold to
suppress the false alarm rate. But both strategies involve trade-offs as well.

First, most stimuli in natural settings are present over periods of time
longer than a few tens of milliseconds. Thus, in rapid coding a miss can be
corrected: for a miss rate $P_{\text{miss}}<1$ in a fundamental time window of
20 ms, a stimulus present during a period of 200 ms allows $\sim10$
opportunities of detection. These multiple opportunities of detection reduce
the overall miss rate to roughly $\left(  P_{\text{miss}}\right)  ^{10}$, a
\textit{much} smaller quantity. However, the consequence is that the false
alarm rate, $P_{\text{false alarm}}$ is the short time window, increases to
roughly $10P_{\text{false alarm}}$ (assuming $P_{\text{false alarm}}\ll1$) in
the long time window. This imbalance can be corrected by raising the detection
threshold, $P\left(  T\mid r\right)  /P\left(  D\mid r\right)  \geq\theta$
(with $\theta>1$ instead of $\theta=1$), so that the false alarm rate goes
down for detection in each fundamental time window. Because the false alarm
rate is suppressed exponentially by raising the threshold, but only increased
linearly by allowing detection in several successive time bins, such a
strategy can be favorable. For instance, in the case of the
\textit{independent} code in Fig. 3, if the threshold is raised to boost the
miss rate to about 10\% (which corresponds to an increase by a factor of 53),
then the false alarm rate is reduced from about 0.1\% down to 0.0001\% (which
corresponds to a suppression by a factor of 850). The obvious cost of this
strategy is that the presence of new objects in the visual world will be noted
slowly, and if there are important objects that require rapid detection this
delay and variability in detection may be unfavorable.

Second, prior expectation can modulate the balance between misses and false
alarms in a favorable manner. The miss rate and the false alarm rate are
weighed by the frequency of occurrence of stimuli, $P\left(  T\right)  $ and
$P\left(  D\right)  $ (see Eqs. (4) and (5) in the Supplementary Experimental
Procedures 1.2). In practice, these quantities are not known and must be
estimated by a freely behaving animal. Changing their values amounts to
weighing the two kinds of error---misses and false alarms---by their
expectation with regards to the occurrence of stimuli. Mathematically, this is
equivalent to weighing miss and false alarm rates as a function of the costs
associated with them. Thus, the effects of expectation and cost can both be
subsumed in the choice of the decoding boundary, $\theta$. If the boundary is
displaced toward the distribution corresponding to Target, then the miss rate
increases while the false alarm rate decreases. The reverse occurs if the
boundary is displaced toward the distribution corresponding to Distracter.
Therefore, an object expected to be incredibly unlikely in a given environment
can have its detection threshold raised substantially to prevent unwanted
false alarms.

This strategy has the obvious drawback that if the rare object \textit{is}
actually present, it will be detected with difficulty. A behaving animal
continually updates its internal representations of expectation and cost as a
function of experience --- a strategy often referred to as Bayesian
decision-making. In a new overall visual context, an otherwise rare object may
be more likely present, and the animal may consequently lower its detection
threshold and, hence, render that object more easily visible. In addition,
temporal integration can enhance the detectability of unexpected objects, thus
helping to overcome a high detection threshold. But of course, both these
methods require more time, so that they will not be effective for rapid
detection. Furthermore, there are limits as to how high the miss rate can be
allowed to increase without adverse behavioral consequences, which places
limits on how effective these strategies can be in achieving very low false
alarm rates.

For all these reasons, it is likely that these strategies are combined with
population codes having intrinsically low error. In fact, the suppression of
the false alarm rate by raising the threshold is much more effective if the
distributions of neural activity are already well separated: in the example of
the \textit{correlated} code in Fig. 3, increasing the miss rate to $10^{-9}$
reduces the false alarm rate by another factor of $10^{15}$.

\section*{Models and Methods}

\subsection*{Definitions of `Macroscopic' and `Microscopic' Correlations}

We consider a neural population divided into $D$ homogeneous pools, labeled by
$\mu,\nu=1,\ldots,D$, and we call $k_{\mu}$ the number of spikes fired in Pool
$\mu$ in a given time bin. The `macroscopic' correlation among pools,
$\chi_{\mu\nu}^{2}$, is defined as%
\begin{equation}
\chi_{\mu\nu}^{2}\equiv\left\langle \left(  k_{\mu}-\left\langle k_{\mu
}\right\rangle \right)  \left(  k_{\nu}-\left\langle k_{\nu}\right\rangle
\right)  \right\rangle .
\end{equation}
The `microscopic' variable which characterizes the state of the neural
population is $s_{i}^{\mu}$; $s_{i}^{\mu}=0$ or $1$depending upon whether the
$i$th neuron in Pool $\mu$ is silent or fires a spike, respectively. The
`microscopic' correlation between neuron $i$ in Pool $\mu$ and neuron $j $ in
Pool $\nu$ is then defined as%
\begin{align}
c_{ij}^{\mu\nu}  &  \equiv\frac{\left\langle \left(  s_{i}^{\mu}-\left\langle
s_{i}^{\mu}\right\rangle \right)  \left(  s_{j}^{\nu}-\left\langle s_{j}^{\nu
}\right\rangle \right)  \right\rangle }{\sqrt{\left\langle \left(  s_{i}^{\mu
}-\left\langle s_{i}^{\mu}\right\rangle \right)  ^{2}\right\rangle }%
\sqrt{\left\langle \left(  s_{j}^{\nu}-\left\langle s_{j}^{\nu}\right\rangle
\right)  ^{2}\right\rangle }}\nonumber\\
&  =\frac{\left\langle \left(  s_{i}^{\mu}-\left\langle s_{i}^{\mu
}\right\rangle \right)  \left(  s_{j}^{\nu}-\left\langle s_{j}^{\nu
}\right\rangle \right)  \right\rangle }{\sqrt{p_{\mu}\left(  1-p_{\mu}\right)
}\sqrt{p_{\nu}\left(  1-p_{\nu}\right)  }},
\end{align}
where $p_{\mu}$ is the firing rate in Pool $\mu$.

Since $k_{\mu}=\sum_{i}s_{i}^{\mu}$, the `macroscopic' correlations are
related to the `microscopic' correlations according to%
\begin{align}
\chi_{\mu\mu}^{2}  &  =\frac{N}{D}p_{\mu}\left(  1-p_{\mu}\right)  \left[
1+\left(  \frac{N}{D}-1\right)  c_{ij}^{\mu\mu}\right]
,\label{micro-macro-relation-1}\\
\chi_{\mu\nu}^{2}  &  =\left(  \frac{N}{D}\right)  ^{2}\sqrt{p_{\mu}\left(
1-p_{\mu}\right)  }\sqrt{p_{\nu}\left(  1-p_{\nu}\right)  }c_{ij}^{\mu\nu},
\label{micro-macro-relation-2}%
\end{align}
where $i\neq j$, $N$ is the total number of neurons in the population and
where we have assumed that all pools have the same size. Hence the identity
between Eqs. (\ref{macro-lockin-condition}) and (\ref{micro-lockin-condition}).

\subsection*{Derivations of the Error Rate}

\textit{Numerical procedure.} The numerical computation of the error rate
relies on maximum entropy probability distributions of the activity of neurons
in each pool. The maximum entropy principle prescribes a unique form of the
probability distribution, which depends on a set of parameters; it is then
assured to be as broad as possible given the constraints on firing rate and
correlation values. The correct parameter values are obtained by the numerical
inversion described in Supplementary Experimental Procedures 1.3. Finally, the
error rate is calculated according to the maximum likelihood rule. See the
Supplementary Experimental Procedures 1.3 for details on maximum entropy and
maximum likelihood prescriptions.

\textit{Analytical procedure.} Here, the relevant variables are taken to be
the spike counts (total number of spikes fired in a given time bin) in each
pool. We then approximate the probability distributions of spike counts by a
Gaussian form; the cross-correlation matrix is related to microscopic
correlations according to Eqs. (\ref{micro-macro-relation-1},
\ref{micro-macro-relation-2}). The error rate is then computed from the volume
of the overlap between pairs of Gaussian probability distributions. See the
Supplementary Experimental Procedures 1.4 for a step-by-step derivation.

\subsection*{Derivations of the Capacity}

In order to maximize the encoding capacity of a neural population given an
arbitrary error rate threshold $\varepsilon^{\ast}$, one has to find the
optimal way of packing probability distribution of response patterns within
the allowed space of spiking responses.

\textit{Independent neurons.} For a pool of independent neurons with average
spike count $k$, the variability in the spike count scales like $\sqrt{k}$.
Thus, probability distributions cannot be arranged in an equidistant manner:
the center-to-center distance between probability distributions must grow at
least as fast as $\sqrt{k}$ (Fig. 5A). Using this consideration, we can
establish a recursive relation between center-to-center distances between
probability distributions, from which we calculate the optimal number of
response distributions allowed in the space of spike counts. See the
Supplementary Experimental Procedures 1.7 for a step-by-step derivation.

\textit{Correlated neurons.} Close to the lock-in limit, the narrow extent of
the response probability distributions is of order of a few spikes per pool
(Fig. 3C and D). Thus, the corresponding, elongated distributions---which are
perforce `parallel to each other' since we assume that the correlation values
do not depend upon the stimulus---can be arrange equidistantly and the
center-to-center distance between neighboring distribution can be of order of
a few spikes (Fig. 5B). As a result, in a system with $N$ neurons and $D$
pools, as many as $\sim\left(  N/D\right)  ^{D-1}$ response probability
distributions can be packed with negligible error rates. See the Supplementary
Experimental Procedures 1.8 for a more detailed argument and step-by-step derivations.

\bigskip

\bigskip

\section*{Acknowledgments}

We are grateful to M. Kardar and J.-P. Nadal for fruitful discussions.

\bigskip

\bibliographystyle{plain}
\bibliography{References-HighFidelityCoding}

\bigskip%

\begin{figure}
[ptb]
\begin{center}
\includegraphics[
height=4.8127in,
width=4.9009in
]%
{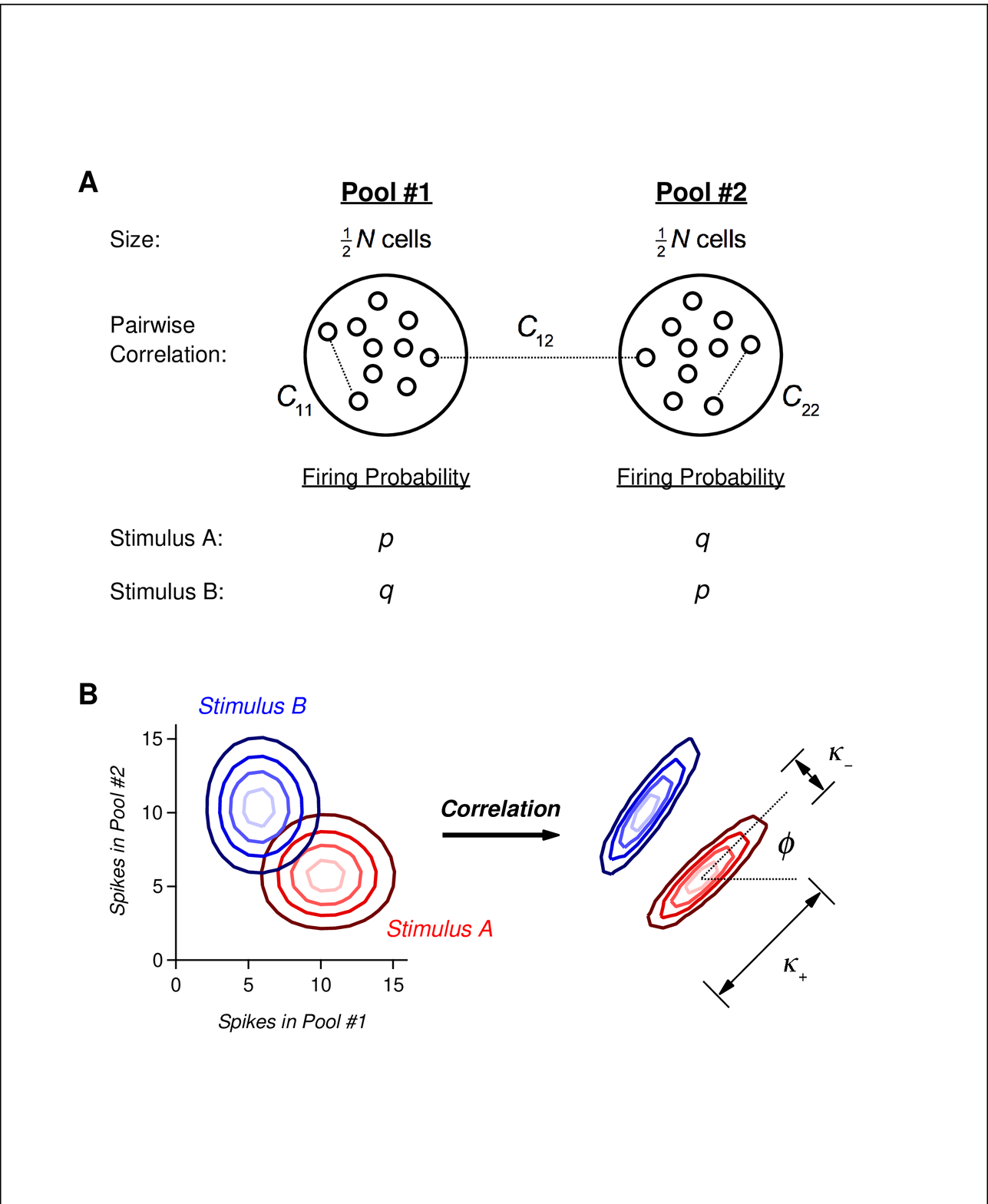}%
\caption{\textbf{Simple model of a population code. A.} Schematics of our
model with two pools with $N/2$ neurons each. Correlation within Pool 1 is
$c_{11}$ for all pairs; correlation within Pool 2 is $c_{22}$ for all pairs;
correlation between the two pools is $c_{12}$ for all pairs. Firing
probability in a single window of time for Pool 1 is $p$ for Target and $q$
for Distracter; firing probabilities are the opposite for Pool 2. \textbf{B.}
Probability contours (lightest shade represents highest probability) for
Target stimulus (red) and Distracter (blue) stimuli in the case of independent
neurons (left). Correlation can shrink the distribution along the line
separating them and extend the distribution perpendicular to their separation
(right). Variances along the two principle axes are denoted by $\varkappa_{+}$
and $\varkappa_{-}$; the angle between the long axis and the horizontal line
is denoted by $\phi$. Variances along the axes of Pool 1 and 2 are denoted by
$\chi_{11}$ and $\chi_{22}$, respectively; the variance across Pools 1 and 2
is denoted by $\chi_{12}$.}%
\end{center}
\end{figure}

\bigskip

\bigskip%

\begin{figure}
[ptb]
\begin{center}
\includegraphics[
height=5.6775in,
width=4.6432in
]%
{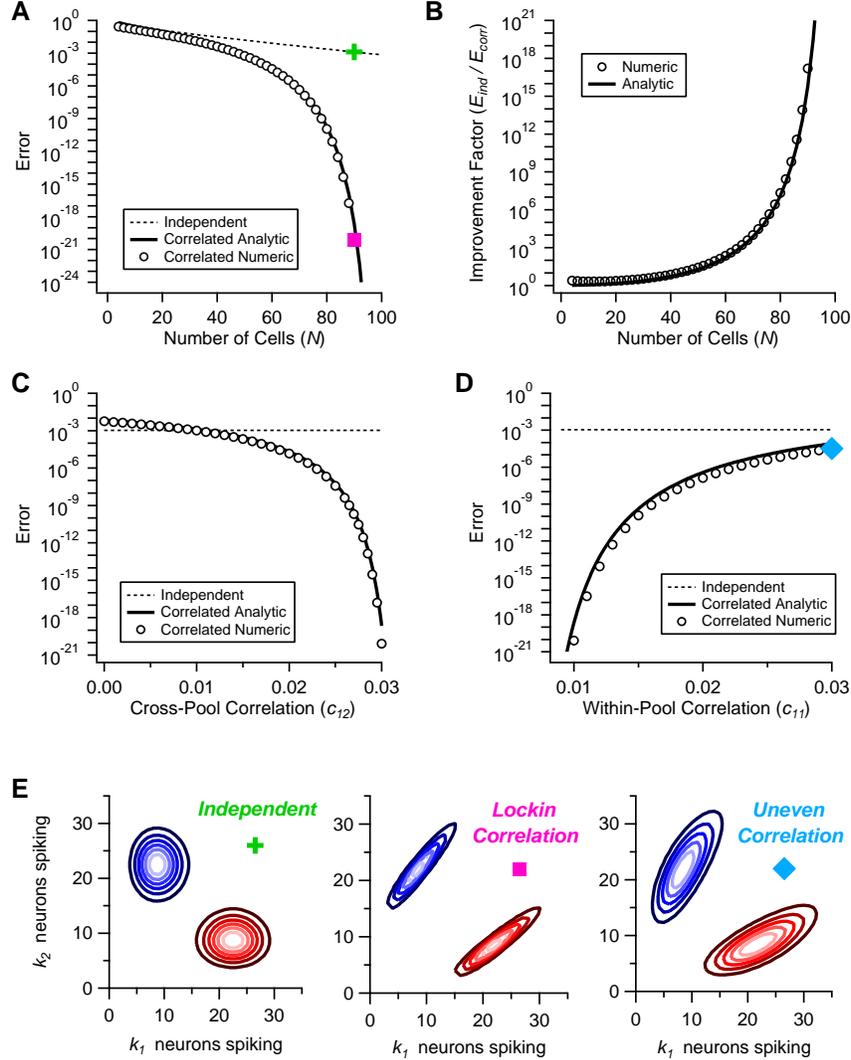}%
\caption{\textbf{Positive correlation can dramatically suppress the error. A.}
Probability of discrimination error for a 2-Pool model of a neural population,
as a function of the number of neurons, $N$, for independent (dashed; all
$c_{ij}=0$) and correlated (circles) populations; parameters are $p=0.5$,
$q=0.2$ for both, and $c_{11}=c_{22}=0.01$, $c_{12}=0.03$ in the correlated
case. Numerical (circles) and analytic (solid line) results are compared.
\textbf{B.} Suppression factor due to correlation, defined as the ratio
between the error probability of independent and correlated populations, as a
function of the number of neurons, $N$; numeric (circles) and analytic (solid
line) results. \textbf{C.} Error probability as a function of the cross-pool
correlation, $c_{12}$, for independent (dashed line) and correlated (circles,
$c_{11}=c_{22}=0.01$) populations; analytic results for correlated population
(solid line). \textbf{D.} Error probability as a function of the correlation
within Pool 1, $c_{11}$, for independent (dashed line) and correlated
(circles, $c_{22}=0.01$, $c_{12}=0.03$) populations; analytic results for
correlated population (solid line). \textbf{E.} Probability contours for three
examples of neural populations; independent (green cross, $N=90$, $p=0.5$,
$q=0.2$), lock-in correlation (pink dot, $c_{11}=c_{22}=0.01 $, $c_{12}%
=0.03$), and uneven correlation (blue diamond, $c_{11}=0.03$, $c_{22}=0.01$,
$c_{12}=0.03$). Colored symbols correspond to points on plots in previous
panels.}%
\end{center}
\end{figure}

\bigskip%

\begin{figure}
[ptb]
\begin{center}
\includegraphics[
height=5.188in,
width=5.4985in
]%
{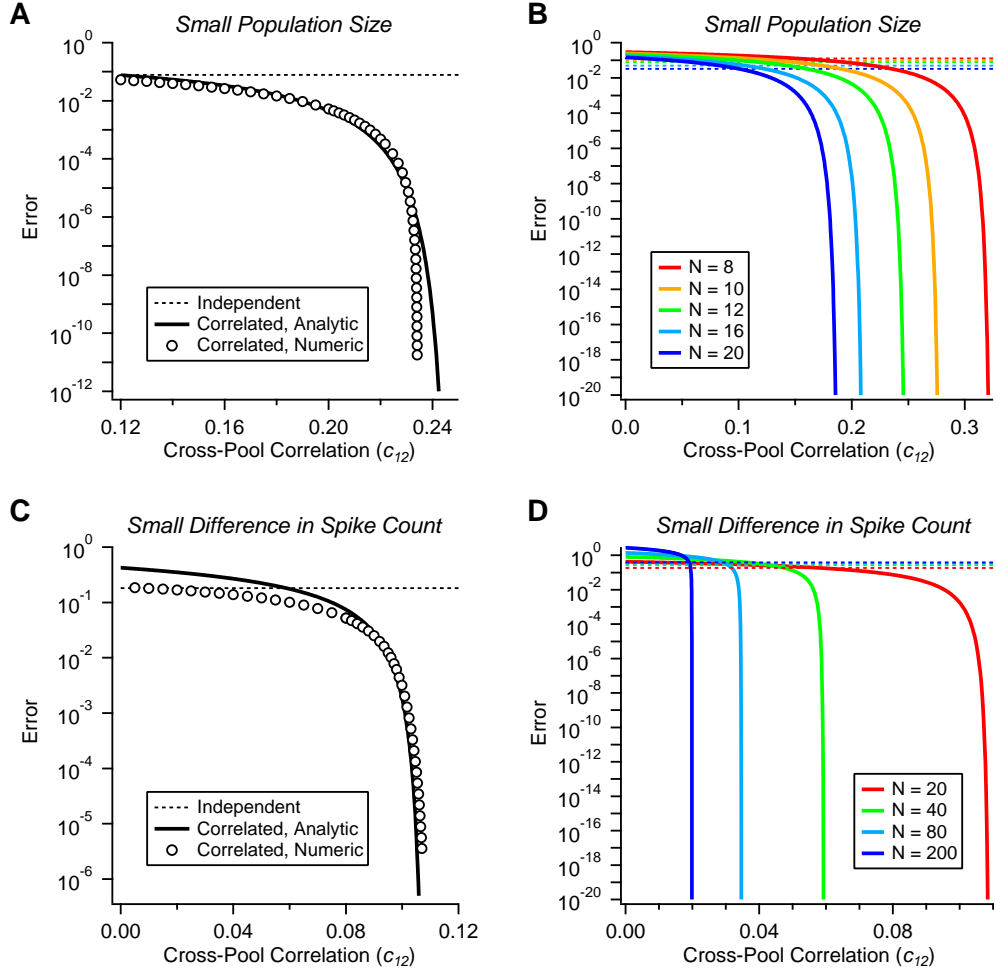}%
\caption{\textbf{Small correlated populations. A.} Probability of error as a
function of the cross-pool correlation, $c_{12}$, for a small neural
population (circles, $N=12$ neurons, $p=0.7$, $q=0.3$, $c_{11}=c_{22}=0.01$),
with analytic result for correlated population (solid line) and independent
population (dashed line) for the sake of comparison. \textbf{B.} Probability
of error \textit{versus} $c_{12}$ for populations of different sizes (colors);
independent population (dashed lines) and analytic results for correlated
population (solid lines). \textbf{C.} Probability of error versus $c_{12}$ for
a neural population with responses differing by an average of 2 spikes ($N=20$
neurons, $p=0.6$, $q=0.4$, $c_{11}=c_{22}=0.01$); numeric solutions (circles),
analytic result (solid line), and independent comparison population (dashed
line). \textbf{D.} Probability of error versus $c_{12}$ for populations having
different sizes but with $N\left(  p-q\right)  $ held constant at 2 spikes
(colors); independent population (dashed lines) and analytic results for
correlated population (solid lines).}%
\end{center}
\end{figure}

\bigskip%

\begin{figure}
[ptb]
\begin{center}
\includegraphics[
height=5.1145in,
width=4.5584in
]%
{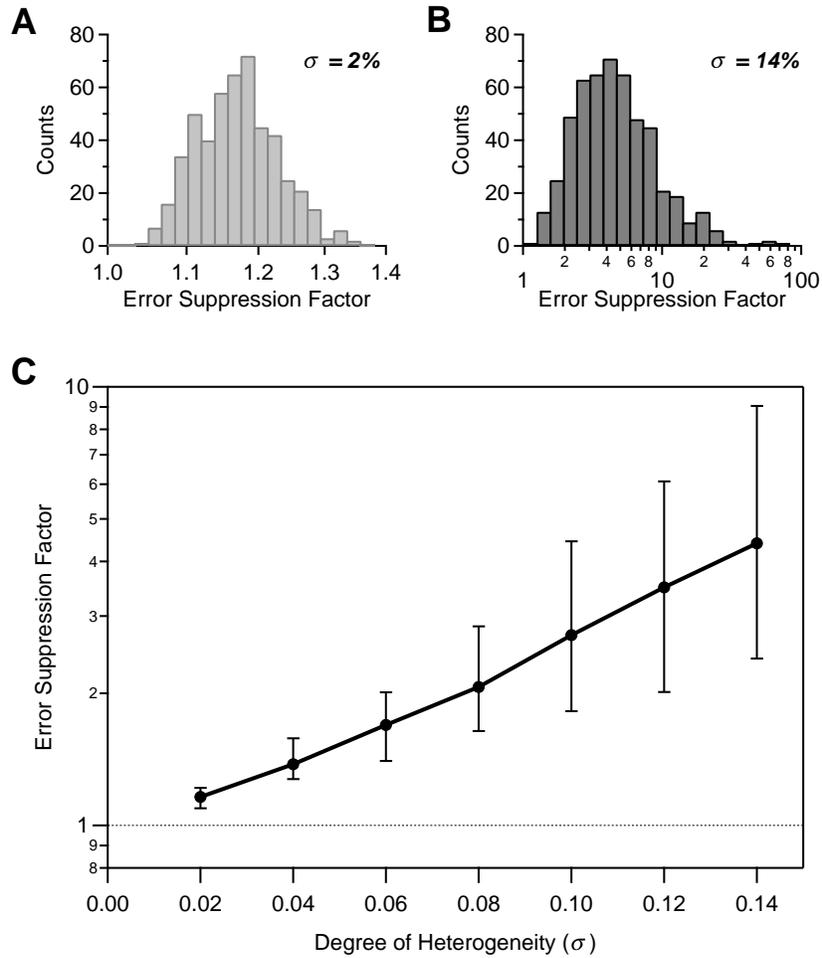}%
\caption{\textbf{Heterogeneous neural populations. A, B.} Histogram of the
error suppression (error in the homogeneous, 2-Pool model divided by the error
in the fully heterogeneous model) for variability values $\sigma=2\%$ and
$14\%$, respectively. All suppression values are greater than one. \textbf{C.}
Value of the error suppression (geometric mean) \textit{versus} the degree of
population variability; $N=10$ neurons, $p=0.7$, $q=0.3$, $c_{11}=c_{22}=0.03
$, $c_{12}=0.21$. (With these parameters, correlation suppresses the error
probability by a factor of 4350 relative to the matched independent
population.)}%
\end{center}
\end{figure}

\bigskip%

\begin{figure}
[ptb]
\begin{center}
\includegraphics[
height=5.361in,
width=5.5798in
]%
{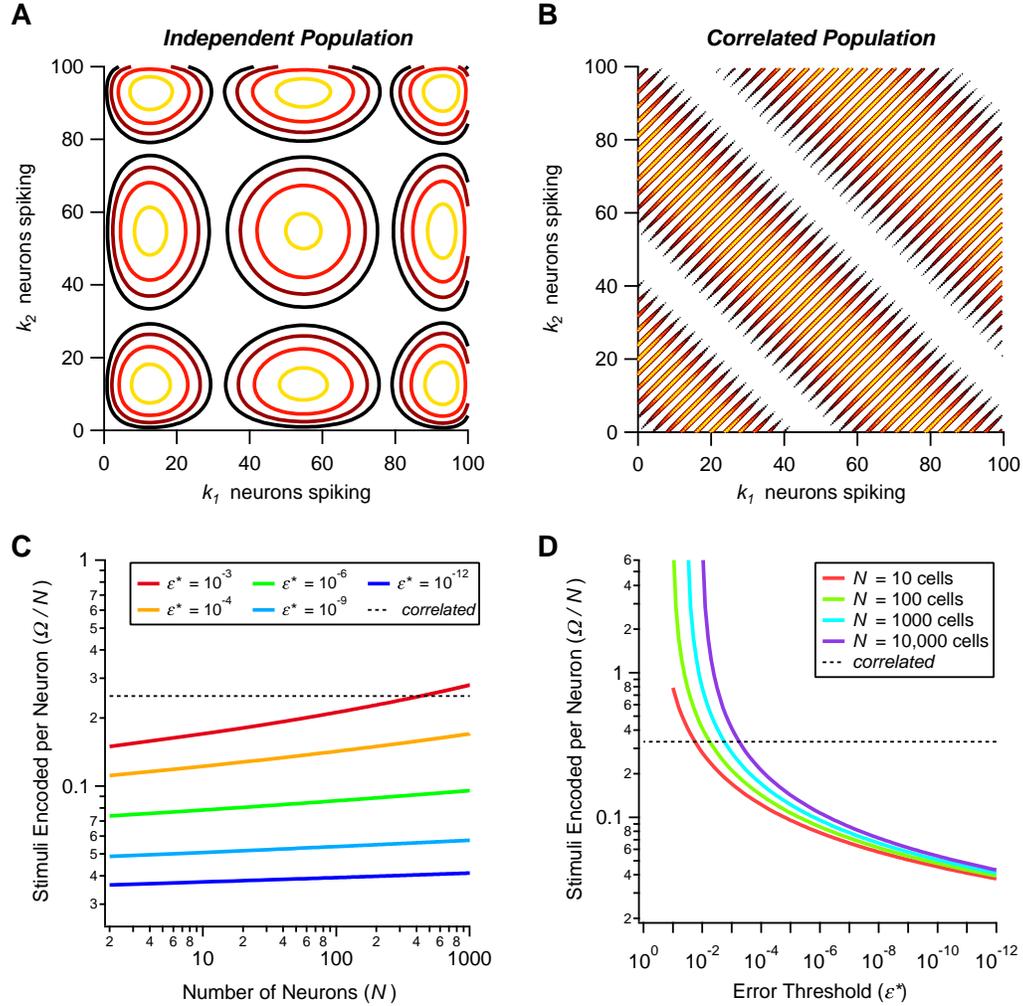}%
\caption{\textbf{Number of encoded stimuli for independent \textit{versus}
correlated populations. A, B.} Schematics of the optimal arrangement of the
probability distributions for independent (A) and correlated (B) populations.
Each set of contours represents the log probability distribution of neural
activity given a stimulus (hotter colors indicate higher probability). Spacing
is set by the criterion that adjacent pairs of distributions have a
discrimination error threshold $\varepsilon^{\ast}=10^{-6}$. \textbf{C.}
Number of stimuli encoded at low error, per neuron, \textit{versus} $N$, for
correlated (dashed line, $a=2$) and independent (solid lines) populations, for
different values of the error criterion, $\varepsilon^{\ast}$ (colors).
\textbf{D.} Number of encoded stimuli per neuron, for correlated (dashed line,
$a=2$) and independent (solid lines) populations, \textit{versus}
$\varepsilon^{\ast}$, for different values of the number of neurons, $N$
(colors).}%
\end{center}
\end{figure}

\bigskip%

\begin{figure}
[ptb]
\begin{center}
\includegraphics[
height=4.8689in,
width=5.5633in
]%
{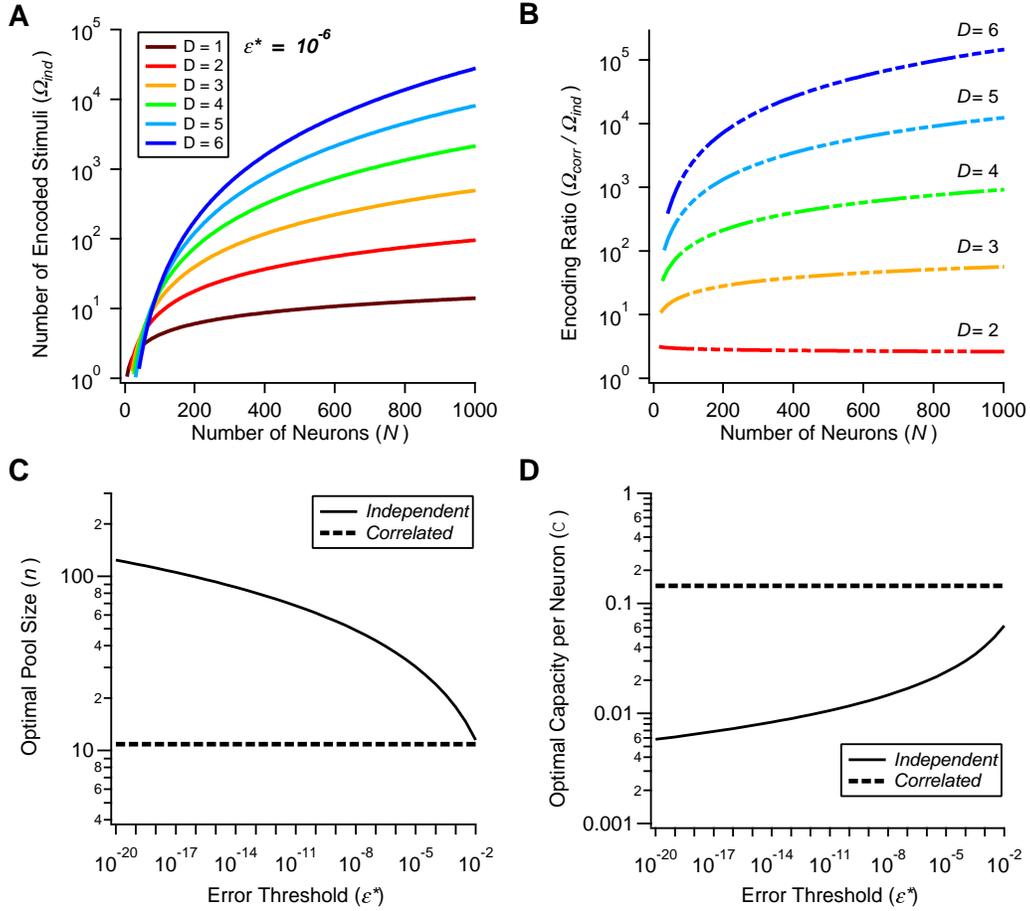}%
\caption{\textbf{Coding capacity of heterogeneous populations. A.} Number of
encoded stimuli \textit{versus} $N$, for an independent population divided
into different numbers of pools, $D$ (colors); the error criterion is
$\varepsilon^{\ast}=10^{-6}$. \textbf{B.} Ratio of the number of encoded
stimuli in a correlated population and the number of encoded stimuli in a
matched independent population, for different numbers of pools $D$ (colors).
\textbf{C.} Optimal pool size, $n$, \textit{versus} error criterion,
$\varepsilon^{\ast}$, for correlated (dashed line, $a=2$) and independent
(solid line) populations. \textbf{D.} Optimal capacity per neuron,
$\mathcal{C}$, \textit{versus} error criterion, $\varepsilon^{\ast}$, for
correlated (dashed line, $a=2$) and independent (solid line) populations.}%
\end{center}
\end{figure}

\bigskip%

\begin{figure}
[ptb]
\begin{center}
\includegraphics[
height=3.9046in,
width=6.2535in
]%
{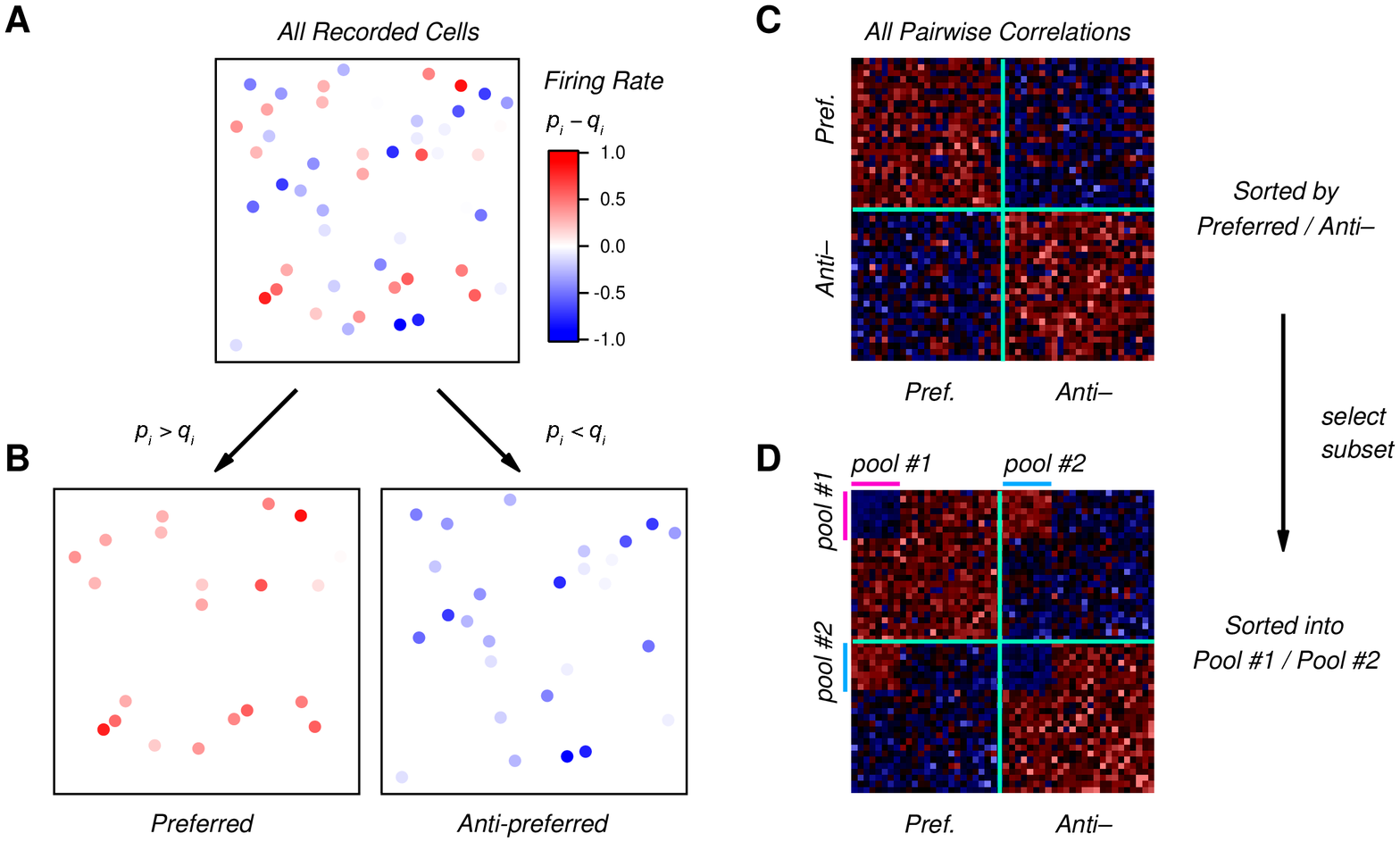}%
\caption{\textbf{Schematics of an experimental test of high-fidelity
correlated coding. A.} Representation of a population of 50 neurons recorded
under two stimulus conditions. Each cell displays firing rates $p_{i}$ and
$q_{i}$ in response to the two stimuli, respectively; the color scale shows
the difference in rates, $p_{i}-q_{i}$. \textbf{B.} The population is divided
into two groups, depending on whether their cells fire more significantly in
response to the first (preferred) or the second (anti-preferred) stimulus.
\textbf{C.} Matrix of correlation values among all pairs of neurons (red =
large, blue = small, black = average), divided into preferred and
anti-preferred groups. Although the overall correlation is stronger for
neurons with the same stimulus tuning (average correlation of pref-pref =
0.206, anti-anti = 0.217, and pref-anti = 0.111), a subset of neurons (Pool 1
and Pool 2) are identified which have the pattern of correlation favorable to
lock-in. \textbf{D.} Matrix of pairwise correlations after re-labeling cells
in order to sort out Pools 1 and 2. Now the favorable pattern of correlation
is visible.}%
\end{center}
\end{figure}

\bigskip\bigskip\newpage

\bigskip

\begin{flushleft}
{\Large \textbf{High-Fidelity Coding with Correlated Neurons}}

\bigskip

\bigskip{\Large Supplementary Material}

{\LARGE \bigskip}

\bigskip

\textsc{Rava Azeredo da Silveira}$^{1,2,3,\ast}$, \textsc{Michael J. Berry
II}$^{4,\dagger}$ \newline

\bigskip

1\textit{\ Department of Physics, Ecole Normale Sup\'{e}rieure, 24 rue
Lhomond, 75005 Paris, France\newline}2\textit{\ Laboratoire de Physique
Statistique, Centre National de la Recherche Scientifique, Universit\'{e}
Pierre et Marie Curie, Universit\'{e} Denis Diderot, France}

3\textit{\ Princeton Neuroscience Institute, Princeton University, Princeton,
New Jersey 08544, U. S. A.\newline}4\textit{\ Department of Molecular Biology,
Princeton University, Princeton, New Jersey 08544, U. S. A. \newline}$\ast
$\textit{\ E-mail: rava@ens.fr \newline}$\dagger$\textit{\ E-mail:
berry@princeton.edu}

\bigskip
\end{flushleft}

\bigskip

\section{Supplementary Methods}

\subsection{Assumptions used in our approach}

The reliability with which the activity of a neural population encodes sensory
inputs can be quantified in a two-alternative forced choice task that
caricatures everyday vision: readout neurons must tell two stimuli apart. In
everyday vision, we often have to identify a given stimulus (\textit{e.g.}, a
known face) against a multitude of other stimuli (\textit{e.g.}, a crowd of
unknown faces), so, by analogy, we refer to the two stimuli in the
discrimination task as `Target' and `Distracter'. These can be two specific
stimuli (\textit{e.g.}, a specific dog \textit{vs.} a specific cat), a
specific stimulus and a stimulus category (\textit{e.g.}, a specific dog
\textit{vs.} all other dogs), or two stimulus categories (\textit{e.g.}, dogs
\textit{vs.} cats); while the book-keeping of input statistics and coding
errors may differ among these cases, the conceptual problem is the same and so
is its quantification. In the companion paper, first we discuss coding
performance in the context of the two-alternative forced choice task. Then we
extend the discussion to the more general case of the discrimination among a
large set of stimuli.

Since we are interested in rapid coding, we focus on short time windows. The
biophysical time scale of neurons---a few tens of milliseconds---affords us
with a natural choice. This time scale also happens to correspond to the spike
timing jitter of individual neurons in the early visual pathway in response to
a natural movie clip \cite{butts_stanley_2007}. We consider short time bins in
which each neuron can only fire one spike or none at all. (As we argue in the
companion paper, this last assumption is not essential. In the more general
case in which many spikes can fit in a time bin, our qualitative conclusions
remain unchanged or may even become stronger.) The situation we have in mind
is one in which a stimulus is presented once every time bin, and the
corresponding population response is recorded. At each occurrence, the
stimulus may be either the Target (which can mean either a specific stimulus
or a class of stimuli), with probability $P\left(  T\right)  $, or the
Distracter (which is all other stimuli and can mean either a specific
stimulus, a stimulus class, or an entire stimulus ensemble), with probability
$P\left(  D\right)  $.

In each time bin, the neural population represents information about the
stimulus in its activity pattern. If the set of population patterns that occur
in response to Target is disjoint from the set of population patterns that
occur in response to Distracter, then the encoded information reaches its
maximum and discrimination can be perfectly reliable. But, quite generally,
neural variability causes some patterns to occur in response to both Target
and Distracter. These cannot be interpreted unambiguously by any deterministic
readout circuit, and some error must result.

\subsection{Maximum likelihood error bound}

In the absence of detailed knowledge about the decoding algorithm employed by
readout neurons, we can still establish a bound on performance. This bound is
derived from maximum likelihood decoding---an algorithm that minimizes the
error rate of deterministic decoding \cite{cover_thomas_1991}. It assigns
Target to a response pattern, $r$, if $P\left(  T\mid r\right)  >P\left(
D\mid r\right)  $ and, conversely, it assigns Distracter to a response
pattern, $r$, if $P\left(  T\mid r\right)  <P\left(  D\mid r\right)  $, where
$P\left(  T\mid r\right)  $ and $P\left(  D\mid r\right)  $ denote the
probability that Target and Distracter, respectively, were presented given
that the response pattern is $r$. The \textit{miss rate}---the fraction of
instances in which Distracter is mistaken for Target---is then calculated as
\begin{equation}
P_{\text{miss}}=\sum_{r\text{ with }P\left(  T\mid r\right)  <P\left(  D\mid
r\right)  }P\left(  T\mid r\right)  P\left(  r\right)  ,
\end{equation}
where $P\left(  r\right)  $ is the probability to record a response pattern
$r$ (regardless of the stimulus presented). Similarly, the \textit{false alarm
rate}---the fraction of instances in which Target is mistaken for
Distracter---is calculated as%
\begin{equation}
P_{\text{false alarm}}=\sum_{r\text{ with }P\left(  D\mid r\right)  <P\left(
T\mid r\right)  }P\left(  D\mid r\right)  P\left(  r\right)  .
\end{equation}
The \textit{total error rate} committed by maximum likelihood decoding,%
\begin{equation}
\varepsilon\equiv P_{\text{miss}}+P_{\text{false alarm}},
\end{equation}
is a lower bound to the error rate committed by any deterministic decoder.
Readout neurons make at least $\varepsilon$ errors per unit time. Throughout,
we use the error rate, $\varepsilon$, as a measure of the fidelity of the
neural population to contrast the coding performance of independent neural
populations \textit{versus} correlated neural populations.

Since experiments record the rate of occurrence of neural responses given the
stimuli, namely the probabilities $P\left(  r\mid T\right)  $ and $P\left(
r\mid D\right)  $, and not the other way around, it is often advantageous to
express the miss and false alarm rates in terms of these measurable
quantities, as%
\begin{equation}
P_{\text{miss}}=\sum_{r\text{ with }P\left(  T\mid r\right)  <P\left(  D\mid
r\right)  }P\left(  r\mid T\right)  P\left(  T\right)  \label{miss-rate}%
\end{equation}
and%
\begin{equation}
P_{\text{false alarm}}=\sum_{r\text{ with }P\left(  D\mid r\right)  <P\left(
T\mid r\right)  }P\left(  r\mid D\right)  P\left(  D\right)  .
\label{false-alarm-rate}%
\end{equation}
In the laboratory, $P\left(  T\right)  $ and $P\left(  D\right)  $ are
controlled by the experimenter; in natural situations, $P\left(  T\right)  $
and $P\left(  D\right)  $ can be thought of as the subject's expectation of
the chances of occurrence of the respective stimuli.

In general misses and false alarms are not symmetric, as they represent
different kinds of errors. In some situations, one may wish to limit the rate
of false alarms more stringently than that of misses, or \textit{vice versa}.
A convenient way to impose such a condition is to introduce a threshold,
$\theta$, greater or smaller than one, when comparing $P\left(  D\mid
r\right)  $ and $P\left(  T\mid r\right)  $, and consequently to generalize
the error rates to%
\begin{equation}
P_{\text{miss}}=\sum_{r\text{ with }\frac{P\left(  T\mid r\right)  }{P\left(
D\mid r\right)  }>\theta}P\left(  r\mid T\right)  P\left(  T\right)  ,
\end{equation}%
\begin{equation}
P_{\text{false alarm}}=\sum_{r\text{ with }\frac{P\left(  T\mid r\right)
}{P\left(  D\mid r\right)  }<\theta}P\left(  r\mid D\right)  P\left(
D\right)  .
\end{equation}
We discuss the asymmetry between misses and false alarms, and the
corresponding role of the threshold, $\theta$, in the Discussion of the
companion paper.

\subsection{2-Pool model of correlated neurons: coding error --- numerical
treatment}

The numerical procedure begins by dividing a population with $N$ neurons into
two homogeneous pools with $N_{1}$ and $N_{2}$ neurons respectively. In terms
of the spike counts in each pool, $k_{1}$ and $k_{2}$, the maximum entropy
distribution \cite{schneidman_berry_2006} of the population activity reads%
\begin{equation}
P\left(  k\mid i\right)  =\frac{N_{1}!}{k_{1}!\left(  N_{1}-k_{1}\right)
!}\frac{N_{2}!}{k_{2}!\left(  N_{2}-k_{2}\right)  !}\frac{\exp\left(  E\left(
k_{1},k_{2}\right)  \right)  }{Z},
\end{equation}
where%
\begin{equation}
E\left(  k_{1},k_{2}\right)  =h_{1}^{\left(  i\right)  }k_{1}+h_{2}^{\left(
i\right)  }k_{2}+\frac{1}{2}J_{11}^{\left(  i\right)  }k_{1}\left(
k_{1}-1\right)  +\frac{1}{2}J_{22}^{\left(  i\right)  }k_{2}\left(
k_{2}-1\right)  +J_{12}^{\left(  i\right)  }k_{1}k_{2}%
\end{equation}
and $i=T$ or $D$. The five parameters, $h_{1}^{\left(  i\right)  }$,
$h_{2}^{\left(  i\right)  }$, $J_{11}^{\left(  i\right)  }$, $J_{22}^{\left(
i\right)  }$, $J_{12}^{\left(  i\right)  }$, are found by direct numerical
solution, such that the firing rates of individual neurons in each other the
two pools, $p$ and $q$, and the normalized pairwise correlations, $c_{11}$
(within Pool 1), $c_{22}$ (within Pool 2), $c_{12}$ (across Pools 1 and 2),
take given values. (Throughout, we borrow symmetric choices (Fig. 2A). That
is, in response to Target the firing rates are $p$ and $q$ in Pools 1 and 2
respectively, while in response to Distracter the firing rates are swapped,
\textit{i.e.}, $q$ and $p$, in Pools 1 and 2 respectively. The same holds for
the correlation values $c_{11}$ and $c_{22}$.) We then apply the maximum
likelihood rule to derive the coding error rate (Figs. 2 and 3).

The choice of maximum entropy distributions is a reasonable one for
establishing upper bounds on the error rate, as these distributions are `as
spread out as possible' given the constraints on firing rates and
correlations. Strictly speaking, true bounds are obtained from minimum mutual
information distributions, but we expect the results to be close to those
obtained from maximum entropy distributions. This expectation is substantiated
by the results obtained from Gaussian approximations---see the remarks at the
end of the next section.

\subsection{2-Pool model of correlated neurons: coding error --- Gaussian
approximation}

We consider a 2-Pool population with $N$ neurons. For the sake of simplicity,
we focus on a symmetric case with $N/2$ neurons in each pool, firing rates $p$
and $q$ in response to Target and Distracter, respectively, in Pool 1, and
vice versa firing rates $q$ and $p$ in response to Target and Distracter,
respectively, in Pool 2. As above, the within-pool correlations are denoted
$c_{11}$ and $c_{22}$, while $c_{12}$ denotes the pairwise correlation across
the two pools. With these hypotheses, a Gaussian approximation to the
probability of response to Target reads
\begin{equation}
P\left(  k_{1},k_{2}\mid T\right)  =\frac{1}{2\pi\sqrt{\text{Det}\chi}}%
\exp\left(  -\frac{1}{2}\left(  K-\left\langle K\right\rangle \right)
\chi^{-1}\left(  K-\left\langle K\right\rangle \right)  ^{T}\right)  ,
\end{equation}
where $k_{1}$ and $k_{2}$ are the spike counts in Pools 1 and 2 respectively.
Here, we use the vector notation with%
\begin{align}
K-\left\langle K\right\rangle  &  =\left(  k_{1}-\left\langle k_{1}%
\right\rangle ,k_{2}-\left\langle k_{2}\right\rangle \right)  ,\\
\left\langle k_{1}\right\rangle  &  =\frac{N}{2}p,\\
\left\langle k_{2}\right\rangle  &  =\frac{N}{2}q,
\end{align}
and the covariance matrix%
\begin{equation}
\chi=\left(
\begin{array}
[c]{ll}%
\chi_{11}^{2} & \chi_{12}^{2}\\
\chi_{12}^{2} & \chi_{22}^{2}%
\end{array}
\right)  .
\end{equation}
A similar expression approximates the probability of response to Distracter,
but with $p$ and $q$ swapped. (The firing rates depend upon the stimulus, but
the correlations do not.) For calculational ease, we give a name to the
inverse covariance matrix:%
\begin{equation}
\chi^{-1}\equiv G\equiv\left(
\begin{array}
[c]{ll}%
g_{11} & g_{12}\\
g_{12} & g_{22}%
\end{array}
\right)  .
\end{equation}

We calculate the maximum likelihood error by saddle-point integration about
the (symmetric) point of maximum equiprobability, $K^{\ast}=\left(  k^{\ast
},k^{\ast}\right)  $, obtained by maximizing the quantity%
\begin{equation}
\left(  K^{\ast}-\left\langle K\right\rangle \right)  \chi^{-1}\left(
K^{\ast}-\left\langle K\right\rangle \right)  ^{T},
\end{equation}
as%
\begin{equation}
k^{\ast}=\frac{\left(  g_{11}+g_{12}\right)  \left\langle k_{1}\right\rangle
+\left(  g_{22}+g_{12}\right)  \left\langle k_{2}\right\rangle }%
{g_{11}+2g_{12}+g_{22}}.
\end{equation}
Then, saddle-point integration yields a coding error rate%
\begin{equation}
\varepsilon\approx\text{Prefactor}\times\exp\left(  -\frac{1}{2}\left(
K^{\ast}-\left\langle K\right\rangle \right)  \chi^{-1}\left(  K^{\ast
}-\left\langle K\right\rangle \right)  ^{T}\right)  .
\end{equation}
Performing the two-dimensional Gaussian integrals and including contributions
from both misses and false alarms, we obtain%
\begin{equation}
\text{Prefactor}\approx\sqrt{\frac{g_{11}+2g_{12}+g_{22}}{\pi\left(
g_{11}g_{22}-g_{12}^{2}\right)  \left(  \left\langle k_{1}\right\rangle
-\left\langle k_{2}\right\rangle \right)  ^{2}}}.
\end{equation}
After we replace $\left\langle k_{1}\right\rangle $ and $\left\langle
k_{2}\right\rangle $ by $Np/2$ and $Nq/2$ respectively, as well as the matrix
elements, $g_{ij}$, by their expression in terms of the pairwise correlation
values, $c_{ij}$, we retrieve Eqs. (1) and (2) in the companion paper. From
the form of Eq. (2), in general we require%
\begin{equation}
c_{12}>\frac{1}{2}\left[  \sqrt{\frac{p\left(  1-p\right)  }{q\left(
1-q\right)  }}c_{11}+\sqrt{\frac{q\left(  1-q\right)  }{p\left(  1-p\right)
}}c_{22}\right]
\end{equation}
for the quantity $\Delta$ to approach zero at large $N$ and, consequently, for
the error rate to become negligible according to Eq. (1). Similarly, at fixed
$N$, increasing the cross-pool correlation $c_{12}$ decreases the denominator
and hence the error rate, while increasing within-pool correlation, either
$c_{11}$ or $c_{22}$, increases the error.

We emphasize the agreement between the numerical and the analytical results
(dots versus solid lines in Figs. 2A-D and Figs. 3A and C), which is not to be
expected in general and is encouraging here. Indeed, numerical results are
derived by making use of maximum entropy distributions. These are as broad as
the constraints on firing rates of individual neurons and pairwise
correlations allow, yet when expressed in terms of spike counts their tails
fall off more rapidly than Gaussian tails. Estimations of the error rate from
maximum entropy distributions and from Gaussian distributions coincide. We
recall that the maximum likelihood error is dominated by the height of the
distributions at equiprobability. So the quantitative similarity between
numerical and Gaussian results means that, even for very stringent error
thresholds, the asymptotic behavior of the tails does not play a dominant role.

\subsection{Arguments for lock-in beyond a Gaussian approximation}

Here, we present general arguments on the role of correlation in high-fidelity
coding, which do not rely on a Gaussian approximation of probability
distributions. We assume only that the probability distributions of spike
counts in response to Target and Distracter are `well-behaved'; specifically,
that they each have a single maximum and that their tails decay rapidly
enough. Then the knowledge of the correlation structure is sufficient to
discuss the degree of their overlap and, hence, the coding error rate. For the
sake of simplicity we still consider a 2-Pool model, but our arguments can be
transposed to the general case of a $D$-Pool model.

We start by examining the quantity%
\begin{equation}
V\left(  \theta\right)  \equiv\left\langle \left[  e\left(  \theta\right)
\cdot\left(  K-\left\langle K\right\rangle \right)  \right]  ^{2}\right\rangle
,
\end{equation}
where $e\left(  \theta\right)  $ is a unit vector along the direction given by
the angle $\theta$ and $K$ is the vector of spike counts. This quantity is
calculated as%
\begin{align}
V\left(  \theta\right)   &  =\left\langle \left[  \cos\left(  \theta\right)
\left(  k_{1}-\left\langle k_{1}\right\rangle \right)  +\sin\left(
\theta\right)  \left(  k_{2}-\left\langle k_{2}\right\rangle \right)  \right]
^{2}\right\rangle \\
&  =\cos\left(  \theta\right)  ^{2}\chi_{11}^{2}+2\sin\left(  \theta\right)
\cos\left(  \theta\right)  \chi_{12}^{2}+\sin\left(  \theta\right)  ^{2}%
\chi_{22}^{2},
\end{align}
\textit{i.e.}, it is the variance along the direction prescribed by the unit
vector $e\left(  \theta\right)  =\left(  \cos\left(  \theta\right)
,\sin\left(  \theta\right)  \right)  $ in the $\left(  k_{1},k_{2}\right)
$-plane of spike counts. Optimizing $V\left(  \theta\right)  $ with respect to
the rotation angle, we find that it reaches its minimal and maximal values,%
\begin{equation}
\varkappa_{\pm}^{2}\equiv\frac{1}{2}\left(  \chi_{11}^{2}+\chi_{22}^{2}%
\pm\sqrt{\left(  \chi_{11}^{2}-\chi_{22}^{2}\right)  ^{2}+4\chi_{12}^{4}%
}\right)  ,
\end{equation}
along the two orthogonal angles given by%
\begin{equation}
\tan\left(  2\theta\right)  =\frac{2\chi_{12}^{2}}{\chi_{11}^{2}-\chi_{22}%
^{2}}. \label{min-max-angles}%
\end{equation}
This expression can also be written in terms of the microscopic correlations
as%
\begin{equation}
\tan\left(  2\theta\right)  =\frac{N\sqrt{p_{1}\left(  1-p_{1}\right)  }%
\sqrt{p_{2}\left(  1-p_{2}\right)  }c_{12}}{p_{1}\left(  1-p_{1}\right)
\left[  1+\left(  N/2-1\right)  c_{11}\right]  -p_{2}\left(  1-p_{2}\right)
\left[  1+\left(  N/2-1\right)  c_{22}\right]  }.
\end{equation}
As mentioned in the companion paper, for positive correlation the angle along
which the distribution elongates, $\theta_{+}=\phi$ (Fig. 1B), lies between
$0^{\circ}$ and $90^{\circ}$. The other solution of this equation lies at
right angle with $\phi$, $\theta_{-}=\phi+90^{\circ}$, and defines the
direction of `probability compression'. The quantities that govern overlap
suppression are the small variances, $\varkappa_{-}$, and the angles $\phi$,
for each of the two distributions corresponding to Target and Distracter. The
error rates decrease with smaller $\varkappa_{-}$ and more parallel distributions.

The positivity of $V\left(  \theta\right)  $ implies a constraint upon the
values of the macroscopic correlations:
\begin{equation}
\chi_{11}\chi_{22}\geq\chi_{12}^{2}. \label{macro-correlations-bound}%
\end{equation}
In terms of the microscopic correlations, the inequality reads%
\begin{equation}
\left[  1+\left(  \frac{N}{2}-1\right)  c_{11}\right]  \left[  1+\left(
\frac{N}{2}-1\right)  c_{22}\right]  \geq\frac{N^{2}}{4}c_{12}^{2}.
\end{equation}
This condition amounts to the positivity of probability. When equality is
achieved, the corresponding probability distribution becomes infinitely thin
along one direction, \textit{i.e.}, the probability of any state in the
$\left(  k_{1},k_{2}\right)  $-plane away from this line vanishes. When
equality is achieved, we say that the neural population is `\textit{locked-in}%
'; in this case, the coding error rate can vanish. When correlation values are
such that the inequality is satisfied, and hence the coding error rate can be
massively suppressed, we refer to the pattern of correlation as
`\textit{favorable}'.

We note in passing that a vanishing error in the Gaussian approximation,
\textit{i.e.}, $\Delta=0$ (see Eq. (2) in the companion paper), corresponds to
two `infinitely thin' probability distributions whose directions of largest
variance are parallel. Indeed, the condition $\varkappa_{-}=0$, which occurs
when $\chi_{11}\chi_{22}=\chi_{12}^{2}$, together with the condition
$\theta=45%
{{}^\circ}%
$ (see Eq. (\ref{min-max-angles}) above) imply%
\begin{align}
1+\left(  \frac{N}{2}-1\right)  c_{11}  &  =\frac{N}{2}\sqrt{\frac{q\left(
1-q\right)  }{p\left(  1-p\right)  }}c_{12},\\
1+\left(  \frac{N}{2}-1\right)  c_{22}  &  =\frac{N}{2}\sqrt{\frac{p\left(
1-p\right)  }{q\left(  1-q\right)  }}c_{12},
\end{align}
\textit{i.e.}, $\Delta=0$.

\subsection{Robustness of high-fidelity coding with respect to parameter
variations}

High-fidelity coding results from the suppression of overlap among response
probability distributions corresponding to different stimuli. By tuning one
combination of the correlation parameters, distributions become thin
(\textit{i.e.}, favorable), and we have demonstrated that this can occur for
realistic values of the correlations. But even in the singular limit of
infinitely thin (\textit{i.e.}, locked-in) distributions, independent
parameters are left free, namely, the orientations of the principal axes of
the distributions or, equivalently, the angles along which the elongated
distributions lie in the $(k_{1},k_{2})$ plane. We have denoted this angle by
$\phi$ (Fig. 1B). An important question is whether these parameters have to be
fine-tuned for high-fidelity coding. In the present section, we show that no
fine-tuning is necessary: high-fidelity coding operates over a wide range of
parameter choices (Supplementary Fig. S1).

Consider, for example, the dependence of the error rate upon the cross-pool
correlation strength, $c_{12}$, for several choices of the angle $\phi$
(Supplementary Fig. S1A). Clearly, when the two distributions corresponding to
Target and Distracter are elongated along the same direction (here the
diagonal, $\phi=45^{\circ}$, because of our choice of symmetric parameters),
the error rate plunges down to vanishing numbers for appropriate correlation
values. If the two distributions are not parallel, there always remains some
overlap, even if they are infinitely thin. However, this overlap is so small
that, even when the angle differs from the diagonal by as much as $20^{\circ}%
$, the error rate is suppressed by more than ten orders of magnitude
(Supplementary Fig. S1A).

In order to explore the parameter dependence of the error rate, we set a
(small) `error rate threshold', $\varepsilon^{\ast}$, not to be exceeded. The
closer $p$ and $q$ are, \textit{i.e.}, the more similar the mean responses to
Target and the response to Distracter, then the more stringent becomes the
threshold condition, $\varepsilon<\varepsilon^{\ast}$, upon the parameters of
the model. An arbitrary threshold---here, we choose $\varepsilon^{\ast
}=10^{-12}$---defines a corresponding `angle bandwidth': a range of
distribution angles, $\phi$, within which the error rate remains below
threshold (Supplementary Fig. S1B). We selected the value of the error
threshold to be sufficiently low that networks within the angle bandwidth
contribute fewer than a single error per human lifetime. Clearly, the angle
bandwidth depends upon all other model parameters. The closer the firing rates
$p$ and $q$ in response to Target and Distracter respectively, the closer the
two distributions lie and, hence, the more precisely their angle has to be
tuned for error rate suppression. Yet, even when the average activities in the
two pools differ by as little as two to five spikes, the angle bandwidth
remains as large as $10^{\circ}$ to $40^{\circ}$ over a wide range of
correlation values (Supplementary Fig. S1B and C). Thus, error rate
suppression is robust to small parameter variations.

\subsection{$D$-Pool model of independent neurons: coding capacity}

For an estimate of the coding capacity of a population of independent neurons,
we approximate the spike count distribution by a Gaussian with appropriate
mean and variance. In the 1-Pool case with $N$ neurons, this distribution
reads%
\begin{equation}
P_{\left\langle k\right\rangle }\left(  k\right)  =\frac{1}{2\pi
\sqrt{\left\langle k\right\rangle \left(  1-\left\langle k\right\rangle
/N\right)  }}\exp\left(  -\frac{1}{2}\frac{\left(  k-\left\langle
k\right\rangle \right)  ^{2}}{\left\langle k\right\rangle \left(
1-\left\langle k\right\rangle /N\right)  }\right)  ,
\end{equation}
where $\left\langle k\right\rangle $ is the mean spike count and $\left\langle
k\right\rangle \left(  1-\left\langle k\right\rangle /N\right)  $ the
variance. We then ask, given one such distribution with parameter
$\left\langle k_{1}\right\rangle $, how far away along the $k$-line should a
distribution, with parameter $\left\langle k_{2}\right\rangle $, be placed so
that the probability not exceed a small value, $\varepsilon^{\ast}$, a the
point of equiprobability, $k^{\ast}$:
\begin{equation}
P_{\left\langle k\right\rangle }\left(  k^{\ast}\right)  =P_{\left\langle
k^{\prime}\right\rangle }\left(  k^{\ast}\right)  \leq\varepsilon^{\ast}.
\end{equation}
We find that we have to require%
\begin{equation}
\sqrt{\left\langle k^{\prime}\right\rangle }-\sqrt{\left\langle k\right\rangle
}\geq\frac{1}{2}\ln\left(  \frac{2}{\pi N\varepsilon^{\ast2}}\right)  ,
\end{equation}
where we have assumed that $\pi N\varepsilon^{\ast2}<2$ and $k+k^{\prime}<N$.
From this recursion, we find that we can `fit' as many as%
\begin{equation}
\Omega_{1\text{-Pool}}^{\text{independent}}\lesssim\sqrt{\frac{4N}{\ln\left(
2/\pi N\varepsilon^{\ast2}\right)  }}%
\end{equation}
well-separated different probability distributions of spike counts, in a
1-pool population with $N$ neurons.

In the 2-Pool case, we calculate similarly the number, $\Omega_{2\text{-Pool
}}^{\text{independent}}$, of well-separated probability distributions that can
be fit within the positive quadrant of the $\left(  k_{1},k_{2}\right)
$-plane of spike counts. Here, $k_{1}$ and $k_{2}$ each run from $0$ to $N/2$
, so $\Omega_{2\text{-Pool}}^{\text{independent}}$ is roughly evaluated as
\begin{align}
\Omega_{2\text{-Pool}}^{\text{independent}}  &  \lesssim\sqrt{\frac{4\left(
N/2\right)  }{\ln\left(  2/\pi\left(  N/2\right)  \varepsilon^{\ast2}\right)
}}\times\sqrt{\frac{4\left(  N/2\right)  }{\ln\left(  2/\pi\left(  N/2\right)
\varepsilon^{\ast2}\right)  }}\nonumber\\
&  =\frac{2N}{\ln\left(  4/\pi N\varepsilon^{\ast2}\right)  }.
\end{align}
We can extend the derivation to the general $D$-Pool case, in which each axis
of the response space runs from $0$ to $N/D$, so that%
\begin{align}
\Omega_{D\text{-Pool}}^{\text{independent}}  &  \lesssim\left[  \sqrt
{\frac{4\left(  N/D\right)  }{\ln\left(  2/\pi\left(  N/D\right)
\varepsilon^{\ast2}\right)  }}\right]  ^{D}\nonumber\\
&  =\left[  \frac{4N}{D\ln\left(  2D/\pi N\varepsilon^{\ast2}\right)
}\right]  ^{D/2}.
\end{align}

By analogy with a population of $N$ deterministic neurons, we define the
\textit{capacity per neuron}, $\mathcal{C}_{D\text{-Pool}}^{\text{independent}%
}$, as%
\begin{equation}
\mathcal{C}_{D\text{-Pool}}^{\text{independent}}\equiv\frac{\log_{2}\left(
\Omega_{D\text{-Pool}}^{\text{independent}}\right)  }{N}.
\end{equation}
In the deterministic case, the population as a whole codes for $2^{N}$ states
and the capacity per neuron is equal to 1 bit. In the case of independent, but
stochastic, neurons,%
\begin{equation}
\mathcal{C}_{D\text{-Pool}}^{\text{independent}}\lesssim\frac{1}{2\ln\left(
2\right)  n}\ln\left(  \frac{2n}{\ln\left(  2/\pi n\varepsilon^{\ast2}\right)
}\right)  ,
\end{equation}
where%
\begin{equation}
n\equiv\frac{N}{D}%
\end{equation}
is the number of neurons per pool. The capacity decreases with decreasing
$\varepsilon^{\ast}$. For a given value of $\varepsilon^{\ast}$, the capacity
is maximal for a characteristic pool size which depends upon $\varepsilon
^{\ast}$ but does not depend upon $N$ and which can be calculated
perturbatively. A rough estimation of this characteristic pool size yields%
\begin{equation}
n_{\text{optimal}}^{\text{independent}}\approx e\ln\left(  \frac{2}{\sqrt{\pi
e}\varepsilon^{\ast}}\right)
\end{equation}
and a maximal capacity per neuron given by%
\begin{equation}
\mathcal{C}_{D\text{-Pool, optimal}}^{\text{independent}}\lesssim\left[
e\ln\left(  2\right)  \ln\left(  \frac{4}{\pi e\varepsilon^{\ast2}}\right)
\right]  ^{-1}. \label{capacity-independent-optimal}%
\end{equation}
Equivalently, the number of stimuli that a population of $N$ independent
neurons can encode with an error threshold $\varepsilon^{\ast}$ is limited\ by%
\begin{equation}
\Omega_{D\text{-Pool, optimal}}^{\text{independent}}\lesssim\exp\left(
\frac{N}{e\ln\left(  4/\pi e\varepsilon^{\ast2}\right)  }\right)  .
\end{equation}

\subsection{$D$-Pool model of correlated neurons: coding capacity}

We derive an estimate of the capacity in the correlated case by evaluating how
many `thin probability distributions' can be fitted in the quadrant of
possible response patterns defined by $0\leq k_{1},k_{2},\ldots,k_{D}\leq
N/D$. In a 2-Pool population ($D=2$), we can arrange one row of `parallel
distributions' along the diagonal that connects the points $\left(
0,N/2\right)  $ and $\left(  N/2,0\right)  $ in the $\left(  k_{1}%
,k_{2}\right)  $ plane. (Three such rows are displayed in Fig. 6B.) If
neighboring distribution centers differ by $O\left(  a\right)  $\ spike, this
manipulation yields a number%
\begin{equation}
\Omega_{\text{2-Pool}}^{\text{correlated}}\approx\frac{N}{2a}%
\end{equation}
of well separated probability distributions that the population can code for.
Similarly, in the general $D$-Pool case we arrange a set of correlated
distributions across a hyperplane within the hypercube with edge $N/D$\ in the
$\left(  k_{1},\ldots,k_{D}\right)  $\ space. Such a configuration immediately
yields a scaling%
\begin{equation}
\Omega_{D\text{-Pool}}^{\text{correlated}}\sim\left(  \frac{N}{D}\right)
^{D-1}=n^{N/n-1},
\end{equation}
where%
\begin{equation}
n\equiv\frac{N}{D}%
\end{equation}
is the number of neurons per pool, as before. To be more precise, we can bound
$\Omega_{D\text{-Pool}}^{\text{correlated}}$\ from below. If we are concerned
that distributions may overlap near the faces of the hypercube, we can, for
example, allow them to fill only a central half of the hyperplane.
Furthermore, if neighboring distribution centers are separated by $a$\ spikes,
we obtain%
\begin{equation}
\Omega_{D\text{-Pool}}^{\text{correlated}}\gtrsim\left(  \frac{N}{2aD}\right)
^{D-1}=\left(  \frac{n}{2a}\right)  ^{N/n-1}.
\end{equation}

This quantity behaves differently from its counterpart in the independent
case: for a wide range of even vanishingly small error thresholds,
$\Omega_{D\text{-Pool}}^{\text{correlated}}$\ is essentially independent of
the error threshold as realistic values of the correlation coefficients can be
chosen so as to make the distributions much narrower than $a$. For fixed $n$,
this bound scales with $N$\ in a trivial manner akin to the independent case.
Indeed, the capacity per neurons,%
\begin{equation}
\mathcal{C}_{D\text{-Pool}}^{\text{correlated}}\equiv\frac{\log_{2}\left(
\Omega_{D\text{-Pool}}^{\text{correlated}}\right)  }{N},
\end{equation}
here becomes%
\begin{equation}
\mathcal{C}_{D\text{-Pool}}^{\text{correlated}}\gtrsim\left(  \frac{1}%
{n}-\frac{1}{N}\right)  \log_{2}\left(  \frac{n}{2a}\right)  \approx\frac
{1}{n}\log_{2}\left(  \frac{n}{2a}\right)  .
\end{equation}
The capacity per neuron is maximized for%
\begin{equation}
n_{\text{optimal}}\approx2ea,
\end{equation}
where $e=\allowbreak2.\,\allowbreak718\,3\ldots$\ is Euler's constant, and is
evaluated as%
\begin{equation}
\mathcal{C}_{D\text{-Pool, optimal}}^{\text{correlated}}\gtrsim\frac{1}%
{\ln\left(  2\right)  n_{\text{optimal}}}\approx\frac{1}{2\ln\left(  2\right)
ea}. \label{capacity-correlated-optimal}%
\end{equation}
We find%
\begin{equation}
n_{\text{optimal}}\approx5\text{\qquad for\qquad}a\approx1
\end{equation}
and%
\begin{equation}
n_{\text{optimal}}\approx10\text{\qquad for\qquad}a\approx2.
\end{equation}
Correspondingly,%
\begin{equation}
\mathcal{C}_{D\text{-Pool, optimal}}^{\text{correlated}}\gtrsim\frac{1}%
{5\ln\left(  2\right)  }\approx0.28\text{\qquad for\qquad}a\approx1
\end{equation}
and%
\begin{equation}
\mathcal{C}_{D\text{-Pool, optimal}}^{\text{correlated}}\gtrsim\frac{1}%
{5\ln\left(  2\right)  }\approx0.14\text{\qquad for\qquad}a\approx2.
\end{equation}
As opposed to the case of independent neurons, here one does not need to
invoke large values of $n$\ for low-error coding. This is because $n$\ is not
the only parameter from which the system can take advantage to suppress error
rates; for each value of $n$, the correlation coefficients may be tuned to
suppress error rates. We emphasize that the result for optimality, with
$n_{\text{optimal}}\approx5-10$, is self-consistent: low-error coding can
indeed occur with such small pool sizes (see Fig. 3).

We find that, in a correlated population, each neuron can carry as much as
$1/6$\ to $1/3$\ bits of information. This result is to be contrasted with the
absolute maximum of $1$\ bit of information in the case of independent,
deterministic neurons and with the corresponding result for independent,
stochastic neurons, Eq. (\ref{capacity-independent-optimal}). In the
correlated case, the optimal capacity per neuron is fixed, whereas in the
independent case it drops with $\varepsilon^{\ast}$. In particular, from Eqs.
(\ref{capacity-independent-optimal}) and (\ref{capacity-correlated-optimal})
with $a\approx2$, we conclude that individual neurons are more informative in
a correlated population, as compared to an independent population, as soon as
the error rate threshold, $\varepsilon^{\ast}$, falls below $0.1$. Thus, for
any realistically small value of the error rate threshold, correlated
populations are favored.

Taking the 2-Pool model as an example, we note that only for relatively large
values of the parameters (e.g., $N>>1000$ or $\varepsilon^{\ast}\gtrsim
10^{-3}$) does $\Omega_{\text{2-Pool}}^{\text{independent}}$ compare with
$\Omega_{\text{2-Pool}}^{\text{correlated}}$. At relatively low threshold
values ($\varepsilon^{\ast}<10^{-6}$), $\Omega_{\text{2-Pool}}%
^{\text{independent}}$ remains well below $\Omega_{\text{2-Pool}%
}^{\text{correlated}}$ for any reasonable (and even large) value of the
population size (Fig. 5D), as the behavior of $\Omega_{\text{2-Pool}%
}^{\text{independent}}$ is dominated by $\varepsilon^{\ast}$ rather than by
$N$ (Fig. 5D). This behavior obtains because the nearly isotropic tails of the
distributions for independent neurons forbid the presence of more than one or
a few distribution centers within the space of neural responses, if the error
threshold is stringent.

It is worth mentioning that for loose error thresholds $\Omega_{\text{2-Pool}%
}^{\text{independent}}$ may exceed $\Omega_{\text{2-Pool}}^{\text{correlated}%
}$. This results from the fact that independent distributions are arranged on
a two-dimensional grid, whereas correlated distributions, which are compressed
along one direction, are arranged along a line (along the `compressed
direction'). Thus, independent distributions can take advantage of the
$\mathcal{O}\left(  N^{2}\right)  $ possible positions of their centers,
whereas correlated distributions have only $\mathcal{O}\left(  N\right)  $ choices.

\subsection{Estimation of the occurrence of favorable patterns of correlation
in cortical networks}

In order to estimate the probability of occurrence of favorable patterns of
correlation in the brain, we rely upon constraints imposed by the following
experimental results. Overall, similar values of the average strength of noise
correlation have been reported in many cortical areas: $c\approx0.2$ in MT
\cite{zohary_newsome_1994, bair_newsome_2001}, V1 \cite{gawne_richmond_1996,
reich_victor_2001, kohn_smith_2005} (but see Ref. \cite{ecker_tolias_2010} for
a report of smaller values), IT \cite{gawne_richmond_1993}, and M1
\cite{maynard_donoghue_1999}, $c\approx0.15$ in somatosensory cortex
\cite{salinas_romo_2000}.\ We note in passing that a comparable number,
$c\approx0.2$, was recorded for complex spikes from nearby Purkinje cells in
the cerebellum \cite{ozden_wang_2008}. Values of pairwise correlation depend
on the time scale they are measured over \cite{averbeck_lee_2004}, with
somewhat smaller values found in time bins of $1-10 $ ms, $c\approx0.1$
\cite{lang_llinas_1999, schneidman_berry_2006}. What is striking about these
observations is the degree of heterogeneity in measured pairwise correlations.
While many studies have emphasized the average value, the distribution of
values spreads in the range $\left[  -0.1,0.5\right]  $ or even beyond
\cite{bair_newsome_2001, maynard_donoghue_1999, kohn_smith_2005}.

Experimental data are not yet detailed enough to resolve the shape of the
distribution with precision. In our calculation, we assume a Gaussian
distribution of correlation coefficients, with a mean of 0.2 and a standard
deviation of 0.2 \cite{bair_newsome_2001, kohn_smith_2005}, although other
similar assumptions, such as flat distribution in the range $\left[
0,0.4\right]  $, do not change our results appreciably. We consider an overall
local population with $N_{0}$ neurons and look for favorable 2-pool
sub-populations contained within it, with $M$ neurons in each pool. The
estimation is obtained by evaluating the number of possible 2-pool populations
within the overall population, on the one hand, and the probability that the
2-pool population is locked in, on the other hand, and then by comparing the
two quantities.

The number of distinct 2-pool populations within the overall population is
calculated as%
\begin{equation}
\mathcal{N}\equiv\binom{N_{0}}{2M}\frac{1}{2}\binom{2M}{M}=\frac{N_{0}%
!}{\left(  N_{0}-M\right)  !M!}\frac{1}{2}\frac{\left(  2M\right)  !}{\left(
M!\right)  ^{2}}.
\end{equation}
Using Stirling's approximation of the factorial and assuming that $M$ is much
smaller than $N_{0}$, we obtain%
\begin{align}
\mathcal{N}  &  =\frac{1}{2}\exp\left(  2M\left[  \ln\left(  \frac{N_{0}}%
{M}\right)  +1-\frac{2M}{N_{0}}\right]  \right) \\
&  \geq\frac{1}{2}\left(  \frac{N_{0}}{M}\right)  ^{2M}.
\end{align}
As for the probability of a lock-in correlation pattern in the 2-pool
population, we require that within-pool correlations not exceed a given value,
$c_{<}$, and cross-pool correlations be at least another value, $c_{>} $.
Lock-in requires that the cross-pool correlation exceed the in-pool
correlation, so we pick an arbitrary correlation value ($c_{<}$) and we pose
that within-pool correlations lie below this value. In the symmetric case, one
approaches lock-in if cross-pool correlation coefficients are comparable to
$c_{>}\equiv1+\left(  M-1\right)  c_{<}$ (from Eq. (9) in the companion
paper). Thus, we examine the probability of occurrence of a 2-pool system with
all in-pool correlations bounded above by $c_{<}$ and all cross-pool
correlations bounded below by $c_{>}$. This probability is calculated as%
\begin{align}
\mathcal{P}  &  =\left(  \eta_{<}^{\frac{1}{2}M\left(  M-1\right)  }\right)
^{2}\eta_{>}^{M^{2}}\\
&  =\left(  \eta_{<}\eta_{>}\right)  ^{M^{2}}\eta_{<}^{-M},
\end{align}
where $\eta_{<}$ and $\eta_{>}$ are cumulative probabilities of weakly and
strongly correlated pairs of neurons respectively. If the data are fitted with
a Gaussian distribution, $G\left(  c\right)  $, then%
\begin{equation}
\eta_{<}=\int_{-\infty}^{c_{<}}dc\,G\left(  c\right)  \label{eta-weak}%
\end{equation}
and%
\begin{equation}
\eta_{>}=\int_{c_{>}}^{\infty}dc\,G\left(  c\right)  . \label{eta-strong}%
\end{equation}

By comparing the above expressions of $\mathcal{N}$ and $\mathcal{P}$, we
obtain estimates of the critical size of a local population, beyond which
favorable patterns of correlation occur with significant probability, and of
the number of favorable patterns that occur randomly within a large local
population. We find that a favorable pattern of correlation is present with
significant probability provided%
\begin{equation}
N_{0}\geq N_{0}^{\text{critical}}\gtrsim\frac{M}{\eta_{<}^{\frac{M-1}{2}}%
\eta_{>}^{\frac{M}{2}}} \label{critical-N}%
\end{equation}
(Supplementary Fig. S2A). In a sufficiently large local population (much
larger than $N_{0}^{\text{critical}}$), one expects to find a large number,
$\nu$, of 2-Pool systems that are close to lock-in. From our above arguments,
we estimate that this number scales roughly as%
\begin{equation}
\nu\approx\left[  \frac{N_{0}}{M}\eta_{<}^{\frac{M-1}{2}}\eta_{>}^{\frac{M}%
{2}}\right]  ^{2M} \label{number-of-subpopulations}%
\end{equation}
(Supplementary Fig. S2B).

In order to optimize the argument, we can tune $c_{<}$ --- the only free
quantity since we have required $c_{>}=1+\left(  M-1\right)  c_{<}$ --- so as
to maximize the product $\eta_{<}\eta_{>}$ and, hence, maximize $\mathcal{P}$.
But we are not concerned by numerical considerations such as this one, as we
have laid down a coarse argument. First, we have assumed that correlation
values for pairs of neurons are drawn randomly from their distribution. This
is likely not the case in reality, where spatial correlations of pairwise
correlations are to be expected in neural populations. Moreover, because of
the bound in Eq. (\ref{macro-correlations-bound}), the 2-pool pattern we
adopted, with in-pool correlations weaker than $c_{<}$ and cross-pool
correlations stronger than $c_{>}=1+\left(  M-1\right)  c_{<}$, is not
realizable even theoretically. (Similar but more complicated correlation
patterns would fix the problem without changing the essence of the argument.)
Second, the $\mathcal{N}$ 2-pool populations, as we count them, are distinct
but overlapping, and hence the probabilities of their correlation patterns are
not independent. The discrepancy is minor provided the product $\eta_{<}%
\eta_{>}$ be sufficiently small.

Our argument assumes a random arrangement of pairwise correlations. Instead,
the brain could use learning mechanisms to select for such patterns and hence
produce them in far greater numbers. What our simple argument illustrates is
that the observed heterogeneity of pairwise correlations makes it plausible
that lock-in patterns of correlation may be found among subsets of neurons in
the cortex. Finally, this discussion has focused on noise correlation, but it
is important to note that if visual discrimination is between classes of
visual stimuli rather than individual stimuli, then signal correlation within
those classes will also contribute to the total correlation in the neural
population. Signal correlations are typically stronger than noise correlations

\section{Supplementary Discussion}

\subsection{Sensory coding requires extremely low error rates}

Everyday vision occurs in a different regime than that probed in many of the
classic studies in visual psychophysics. Our retina is presented with
complicated scenes in rapid succession---either because of saccadic eye
movements or because of motion in the scene itself---from an enormous set of
possibilities. Often, we seek to recognize the presence of a target stimulus
or stimulus class and distinguish it from every other possible stimulus. For
example, we might want to recognize a friend's face in a particular spatial
location. That location might contain another person's face, or a flower, or
myriad other objects, which we do not want to mistake for our friend's face.
Alternatively, the target stimulus is often a class of related stimuli, such
as that friend's face from a variety of angles or the presence of any human
face, so that a class of visual patterns on the retina, rather than a single
fixed pattern, is to be identified.

In this regime, one distinguishes two kinds of coding error: \textit{misses}
and \textit{false alarms}. In the former, one does not pick up on the target
stimulus; in the latter, an absent target stimulus is erroneously perceived.
While both kinds of error take place occasionally (think of mistaking a wavy
tree branch for a snake, as a false alarm), the effortless feat of the visual
system in avoiding them most of the time is rather bewildering. If we pause a
moment on what this feat means at the neural level, as illustrated by the
following example, we realize that it requires extremely precise coding.

Imagine stretching out on your hotel bed in a tropical country. If there were
a very large spider on the ceiling, you most likely would want to detect it
and detect it promptly. For the sake of concreteness, let us imagine that the
spider has a size of three centimeters and is three meters away, subtending a
visual angle of 0.01 radians. Thus, there are ~$1/(0.01)^{2}=10^{4}$ possible
spider locations on the ceiling. If you are able to detect the spider in any
of these locations, it implies that your brain must effectively have a
`spider-detector' circuit that reads out activity from a retinal population
that subtends each of these spatial locations. If you would like to detect the
spider quickly, say in 100 milliseconds, then there are ~$10^{5}$ possible
spider-detection events per second. Now, if each detector operates at a false
alarm rate that would naively seem low enough to be acceptable, say
0.001---\textit{i.e.}, a probability of error of a tenth of a percent--- you
would still perceive 100 virtual spiders per second!

One can think of a number of resolutions to this `spider-on-the-wall problem'
(changing hotel rooms will not do). Temporal integration, for one, may be used
to suppress errors. Also, error rates ought to be influenced by the prior
expectation of an event---a quantity we have not included explicitly in our
argument. That said, both temporal integration and prior expectation involve
trade-offs. Extensive temporal integration requires longer viewing times, and
many behaviors need to occur quickly. Relying too heavily upon prior
expectation could leave one unable to recognize novel objects.

A more direct way of ensuring reliable discrimination is to employ neural
populations that are organized to suppress false alarm (and miss) rates down
to extremely low values. In the companion paper we focus on this strategy. As
an illustration of the stringency of the requirement, imagine that no more
than one virtual spider ought to be perceived in the hour it takes you to fall
asleep (as such spider detections could prevent sleep). This condition is
satisfied if the false alarm rate remains below $\sim10^{-8}$ per detection
circuit. And of course, the visual system can recognize many objects other
than spiders, implying even lower false alarm rates in any one kind of
detector so that the total false alarm rate remain very low.

\subsection{Arguments for the detrimental effect of positive correlation on
coding with a homogeneous neural population}

It is often said that positive correlation is detrimental to coding. This
claim is based on intuition developed for homogeneous populations
\cite{sompolinsky_shamir_2001, zohary_newsome_1994}, as we explain in this
section. Imagine turning on positive correlation in the population response to
Target (Supplementary Fig. S3A). Distributions of population activity for
increasing values of correlation are progressively wider, causing greater
overlap and hence an enhanced coding error rate. This behavior is generic for
positive values of the correlation (Supplementary Fig. S3B). (In extreme,
non-generic cases with very large values of the correlation, the distribution
corresponding to Target may become bimodal and concentrated around 0 and $N$.
The overlap between the two distributions can then decrease, and hence coding
can improve. But such extreme cases are very different qualitatively from the
experimental situation in which pairwise correlations are small to moderate,
ranging from -0.1 to 0.5 \cite{bair_newsome_2001, fiser_weliky_2004,
kohn_smith_2005, lang_llinas_1999, lee_georgopoulos_1998}. In contrast to
positive values of the correlation, negative values reduce the discrimination
error but, again, such values are rarely observed experimentally.)

Simple arguments explain this behavior. Positive correlations enhance
fluctuations in the population response, as compared to the independent case,
and, as a result, suppress the signal-to-noise ratio. If $r_{i}$ denotes the
response of neuron $i$, the variance of the population activity is%
\begin{equation}
\left\langle \left(  \sum_{i}r_{i}-\left\langle \sum_{i}r_{i}\right\rangle
\right)  ^{2}\right\rangle =\sum_{i}\left\langle \left(  r_{i}-\left\langle
r_{i}\right\rangle \right)  ^{2}\right\rangle +\sum_{i\neq j}\left\langle
\left(  r_{i}-\left\langle r_{i}\right\rangle \right)  \left(  r_{j}%
-\left\langle r_{j}\right\rangle \right)  \right\rangle ,
\end{equation}
where brackets indicate an average over trials. The first sum on the
right-hand-side pertains to fluctuations in single-neuron responses and is
non-vanishing in both independent and correlated cases. The second sum on the
right-hand-side pertains to correlations among neurons. For negative
correlations (anti-correlations), this sum is negative and, hence, the
distribution of neural activity is more narrowly peaked than in the
independent case. By contrast, positive correlations broaden the distribution.
In the anti-correlated case, distributions of population activity
corresponding to different stimuli tend to be well separated, while in the
positively correlated case, overlaps tend to be greater. Therefore,
homogeneous populations with positive correlation have worse coding
performance than corresponding independent populations and, consequently,
require more neurons to achieve a low rate of coding errors.

We can understand the hindrance of coding performance from positive
correlations in an alternate, simple fashion. A homogeneous population with
positive correlation behaves, effectively, as a smaller population. In the
limiting case of a perfectly correlated population in which all neurons
respond identically, the entire population behaves as one, big neuron. Hence,
we expect such positively correlated populations to code information with less
`resolution' and, consequently, to commit coding errors more often than
corresponding independent populations do.

\subsection{High-fidelity coding bare bones}

In the companion paper we demonstrate, quantitatively and with the use of
simple models, that positive correlation can suppress coding errors and
enhance coding capacity massively. The basic mechanism behind this effect was
noted by a number of authors \cite{johnson_1980, vogels_1990,
abbott_dayan_1999, sompolinsky_shamir_2001, averbeck_pouget_2006} and is
simple to understand: positive correlations can deform the shape of
probability distributions of neural activity in such a way as to sharpen the
distinction between nearby probability distributions (Fig. 1B). Put
differently, while positive correlations have a broadening effect overall,
they can nonetheless suppress the tails of probability distributions along
relevant directions, thereby reducing the unfavorable effect of neural variability.

The same idea can be expressed in a more general fashion: the structure of
correlation can be such that it relegates noise into a non-informative mode of
the neural population response. A simple example provides a nice illustration
(\cite{nadal_private}; a similar argument is presented in Ref.
\cite{abbott_dayan_1999}). Consider two neurons with responses
\begin{align}
r_{1}  &  =m_{1}+\delta_{1},\\
r_{2}  &  =m_{2}+\delta_{2}.
\end{align}
We assume that the mean responses, $m_{1}$ and $m_{2}$, are different, such
that $m_{1}$ ($m_{2}$) is large (small) in response to the Target stimulus,
and \textit{vice versa} for the Distracter stimulus. The additive
variabilities, $\delta_{1}$ and $\delta_{2}$, are highly correlated, such that
$\delta_{1}\approx\delta_{2}$. \ Then the informative mode,%
\begin{equation}
r_{-}\equiv r_{1}-r_{2}\approx m_{1}-m_{2},
\end{equation}
is close to noiseless, while all the noise is relegated to the uninformative
mode,%
\begin{equation}
r_{+}\equiv r_{1}+r_{2}\approx m_{1}+m_{2}+\delta_{1}+\delta_{2}.
\end{equation}
Our results can also be viewed in terms of a similar mechanism: informative
and uninformative modes correspond to combinations of pool spike counts, the
$k_{i}$s, and given patterns of positive correlations relegate variability to
the uninformative modes. In the simplest, symmetric, 2-pool model, correlation
sharpens the response distributions along the informative mode, $k_{1}-k_{2}$,
while it blurs them along the uninformative mode, $k_{1}+k_{2}$. Clearly, it
is a signature of correlated coding that informative modes can be identified
only when simultaneous activities of the neurons in the population are considered.

\subsection{How extreme is lock-in?}

In the limit of large enough populations or strong enough pairwise
correlations, the distribution of activity of the population can `lock-in' to
a state of lower dimensionality. We have shown that a neural population can
approach this state in cases in which the pairwise correlations have moderate
values, but we still might wonder how `extreme' is the lock-in state at the
population level? For example, is the population `frozen' at lock-in, with all
variability suppressed? The positivity of probability implies constraints upon
moments of the neural activity; in particular, we have $\chi_{11}\chi_{22}%
\geq\chi_{12}^{2}$. This bound is achieved by the lock-in condition, Eq. (6)
in the companion paper. Thus, lock-in embodies the limiting case of maximum
\textit{macroscopic} correlation between Pools 1 and 2, but there remains a
significant amount of (microscopic) variability even at lock-in. The
specificity of lock-in is that it forbids a subset of the microscopic
patterns, \textit{i.e.}, that these occur with vanishing probability. At
lock-in, the system is not confined to one output pattern. A large set of
patterns can occur with non-negligible probability each---hence the
variability---and the remaining patterns are ruled out---hence the vanishing
overlaps and error rates.

In the Gaussian approximation of the two-pool model, only patterns with a
fixed ratio between $k_{1}-\left\langle k_{1}\right\rangle $ and
$k_{2}-\left\langle k_{2}\right\rangle $ are allowed at lock-in. In the
absence of correlations, allowed output patterns fill a two-dimensional
space---the $\left(  k_{1},k_{2}\right)  $ plane. When correlations push the
system to lock-in, output patterns are confined to a one-dimensional
space---the ($k_{1}-\left\langle k_{1}\right\rangle )\propto(k_{2}%
-\left\langle k_{2}\right\rangle )$. From this dimensionality reduction
results error rate suppression and increased capacity. Of course, the
population attains the actual lock-in state only for specific values of
pairwise correlation and firing rate; however, we have shown that the error
rate can reach near-vanishing values for a range of parameters that do not
bring the population all the way to the lock-in condition. This result is
generic as it relies only upon the rapid fall-off of the tails of the response
probability distribution. It obtains in the case of the Gaussian
approximation, and is bound to apply to the maximum entropy distribution as
the latter's tails are sharper than Gaussian tails.

\subsection{Downstream read-out circuits and decoding}

The issue of how information encoded by neural activity in one brain region
can be read out by other brain regions is an important topic that pertains to
almost all studies of the neural code. It is also a very difficult problem, as
evidenced by the fact that we don't have certain answers to these kinds of
questions in any particular system (at least as far as the authors are aware).
What one can do is to propose decoding algorithms that can read out relevant
information. Such proposals do not imply that read-out brain regions actually
use such algorithms; instead, they are merely existence proofs that
information can be retrieved, and they represent bounds on the performance of
read-out regions.

For instance, in our example of the two-pool model, the decision boundary that
separates response patterns best interpreted as representing Target from those
best interpreted as representing Distracter is a line in the space of neural
responses: $k_{1}=k_{2}$. Thus, a particularly simple linear decoder could
read out the information encoded by the two correlated neural pools at a level
of performance matching the maximum likelihood rule that we used in our
mathematical analysis. In the case of multiple stimuli encoded by two pools,
the decision boundaries generalize to $k_{1}=\alpha k_{2}+\beta$. In addition,
the readout circuit must combine together similar decoding rules for multiple
decision surfaces. Note, however, that this is the same level of complexity as
is needed to read out information in the case of an independent neural code.

In the more realistic case of a fully heterogeneous neural population (as
analyzed in Fig. 4), a decoder that reads out information from the correlated
neural population at optimal performance would need to have the form of a
maximum entropy model \cite{schneidman_berry_2006}, and the decision surface
could be arbitrarily curved in the space of neural responses. Of course, one
might be able to recover nearly optimal performance with simpler decoders if
the heterogeneity is not too severe. This can only be ascertained by further
study of decoding algorithms, and would certainly depend in detail on the
magnitude and pattern of heterogeneity. We have not included any analysis of
decoding mechanisms in this manuscript, because we feel that this is a
substantial topic best left for subsequent studies.

The question of how the brain finds the right neurons form which to extract
relevant information is also important, but unsolved. However, experiments on
brain-machine interfaces demonstrate that the brain has a truly remarkable
ability to change its circuitry in sensory-motor pathways to activate the
relevant motor neurons. In an experiment, $\sim100$\ neurons in primary motor
cortex were recorded and their responses were used to drive movements of
cursors or even robotic arms with simple cosine tuning functions
\cite{velliste_perel_2008}. Under these circumstances, monkeys were able to
achieve high performance in directing movements. Most impressively, the
authors showed that they could re-arrange the tuning curves used to translate
neural activity into movements of the robotic arm, and monkeys could change
their entire sensory-motor pathway in order to fire those particular neurons
in the right pattern to achieve the desired movement
\cite{jarosiewicz_chase_2008}. In many cases, the tuning curves were
completely inverted, and yet the monkeys re-learned how to fire those neurons
appropriately. Analogous results have been reproduced by another lab
\cite{ganguly_carmena_2009}. A related example comes from experiments in which
human subject wear inverting prism glasses. Initially, the world appears
upside-down, resulting in profound motor deficits and disorientation. But
after about a week, subjects regain their coordination, evidently requiring a
complete remapping (inversion) of visual stimuli to motor outputs
\cite{richter_magnusson_2002}.

So, while the brain faces great difficulties in obtaining useful information
encoded by sensory circuits and must be subject to certain limits in
accomplishing these, it is clear that the brain has a remarkable ability to
surmount these difficulties in many situations. We currently have very little
understanding of how the brain manages this, and hence we really don't know at
this point what the limitations are.

Another important issue concerns the manner in which multiple stimuli encoded
by a correlated population are read-out by downstream brain circuits.
Obviously, each stimulus or stimulus class that must be discriminated from all
the others will require a dedicated read-out circuit with specific choices of
synaptic weights for their inputs from the encoding population. Because the
parameters of any single read-out circuit will need tuning in order to achieve
high performance, we have an overall picture of this process in which not all
possible stimulus discriminations are actually read out by subsequent brain regions.

To illustrate this picture, consider the case of human recognition of the
letters of the alphabet. Depending on where a person is born, they learn
different languages, which use different character sets comprising essentially
arbitrary sets of spatial patterns (e.g., Chinese character set versus Latin
or Cyrillic alphabets). At adult levels of performance, these characters can
be discriminated rapidly and with extremely low error. As far as we can tell,
any human child has the ability to learn any language. What this implies is
that the early visual system must encode all possible characters of all human
languages with high fidelity. However, higher centers in the visual pathway
will only develop high-performance readout circuits to process the characters
of languages that a given person actually knows. As the need for other spatial
pattern discrimination arises, new circuits can be learned to read out that
information from early visual areas. This picture is in overall agreement with
the properties of higher visual centers in the ventral stream, where one
observes highly specific feature selectivity, such as to individual faces, and
where feature selectivity depends strongly on individual experience.

\bigskip

\bibliographystyle{plain}
\bibliography{,,,,,,References-HighFidelityCoding}

\bigskip%

\begin{figure}
[ptb]
\begin{center}
\includegraphics[
height=5.6299in,
width=6.3322in
]%
{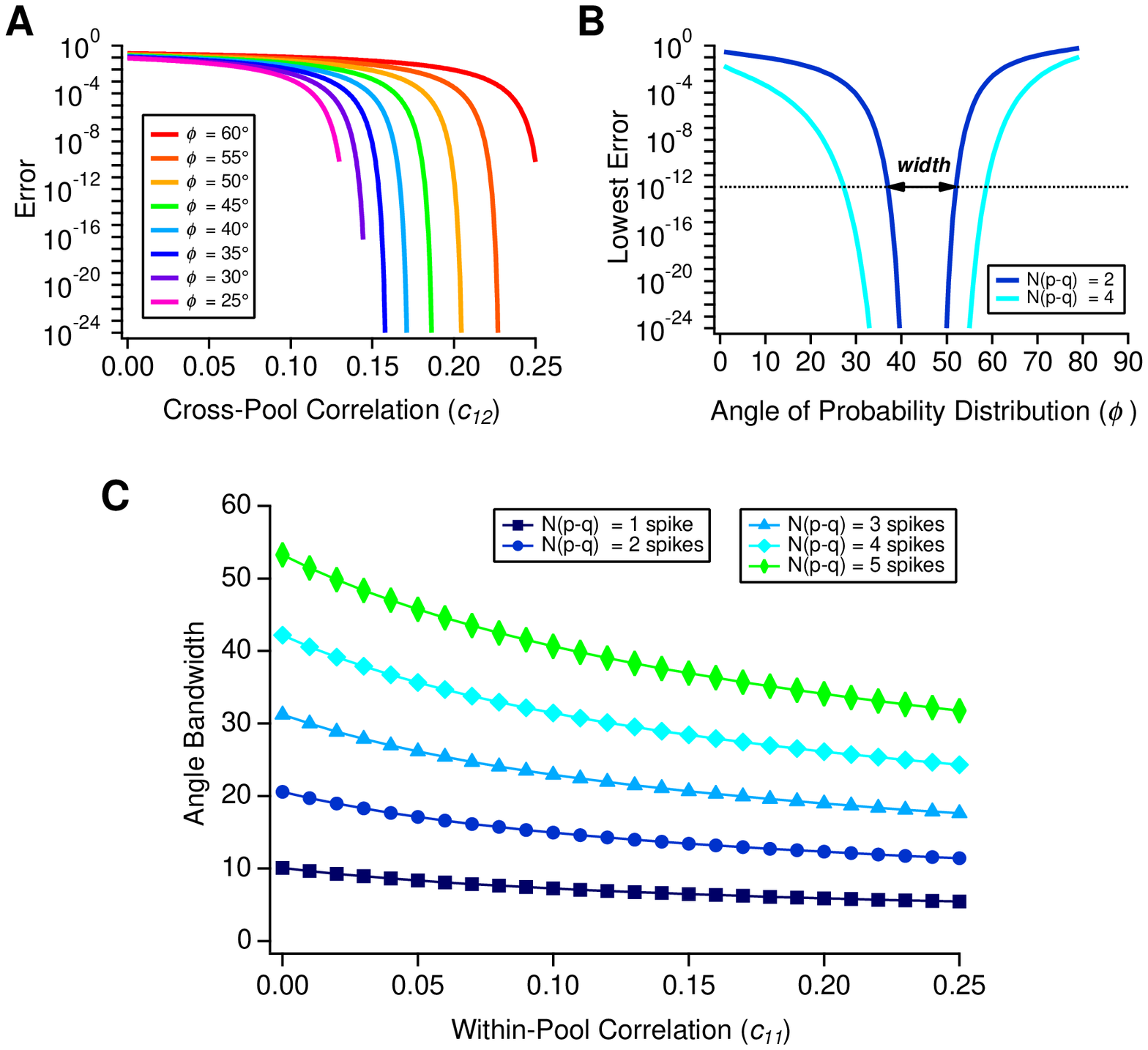}%
\caption{\textbf{(Figure S1.) Robustness to parameter variations. A.}
Probability of error as a function of the cross-pool correlation $c_{12}$ for
populations with $N=20$ neurons and different angles $\varphi$ of their
probability distributions in the space of $(k_{1},k_{2})$ (see Fig. 1 in the
main text); parameters are ($p=0.7$, $q=0.3$, $c_{11}=0.1$) with $c_{22}$ set
to give the chosen angle (Suppl. Eq. (32)). \textbf{B.} Probability of error
as a function of angle for fixed difference in spike count, $N(p-q)$,
intersects the error criterion $\varepsilon^{\ast}=10-12$ at two angles, which
defines the angular bandwidth. \textbf{C.} Angular bandwidth plotted as a
function of within pool correlation, $c_{11}$, for different values of the
difference in spike count, $N(p-q)$.}%
\end{center}
\end{figure}

\bigskip%

\begin{figure}
[ptb]
\begin{center}
\includegraphics[
height=6.5475in,
width=5.297in
]%
{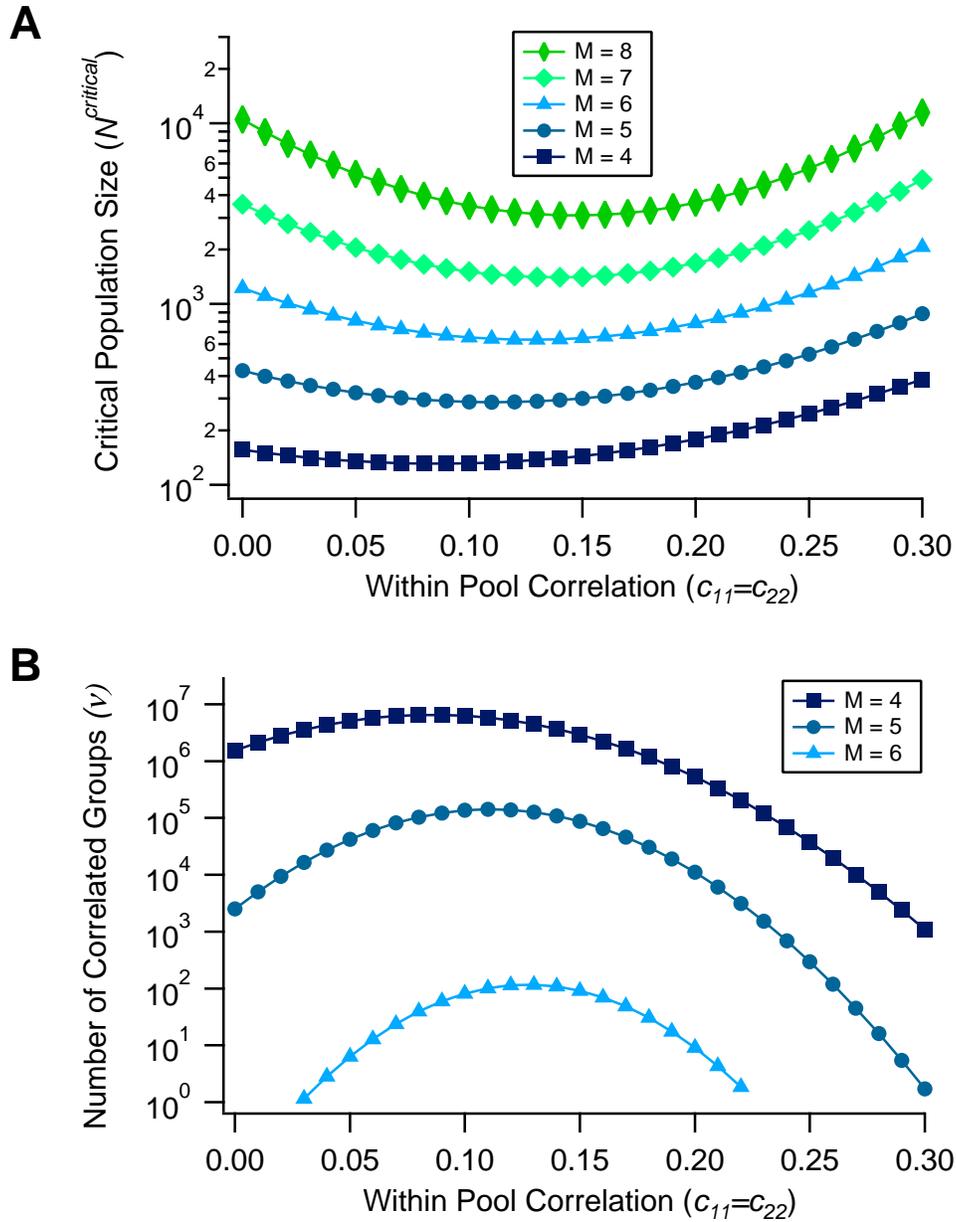}%
\caption{\textbf{(Figure S2.) Lock-in correlations among random populations.
A.} Critical population size $N_{0}^{\text{critical}}$ for randomly finding
groups of neurons with lock-in correlation plotted as a function of within
pool correlation strength ($c_{11}=c_{22}$) for different population sizes
(colors). \textbf{B.} Number of groups of neurons with lock-in correlations in
a local population of $1000 $ neurons plotted as a function of within pool
correlation strength ($c_{11}=c_{22}$) for different population sizes
(colors).}%
\end{center}
\end{figure}

\bigskip%

\begin{figure}
[ptb]
\begin{center}
\includegraphics[
height=5.8064in,
width=4.7608in
]%
{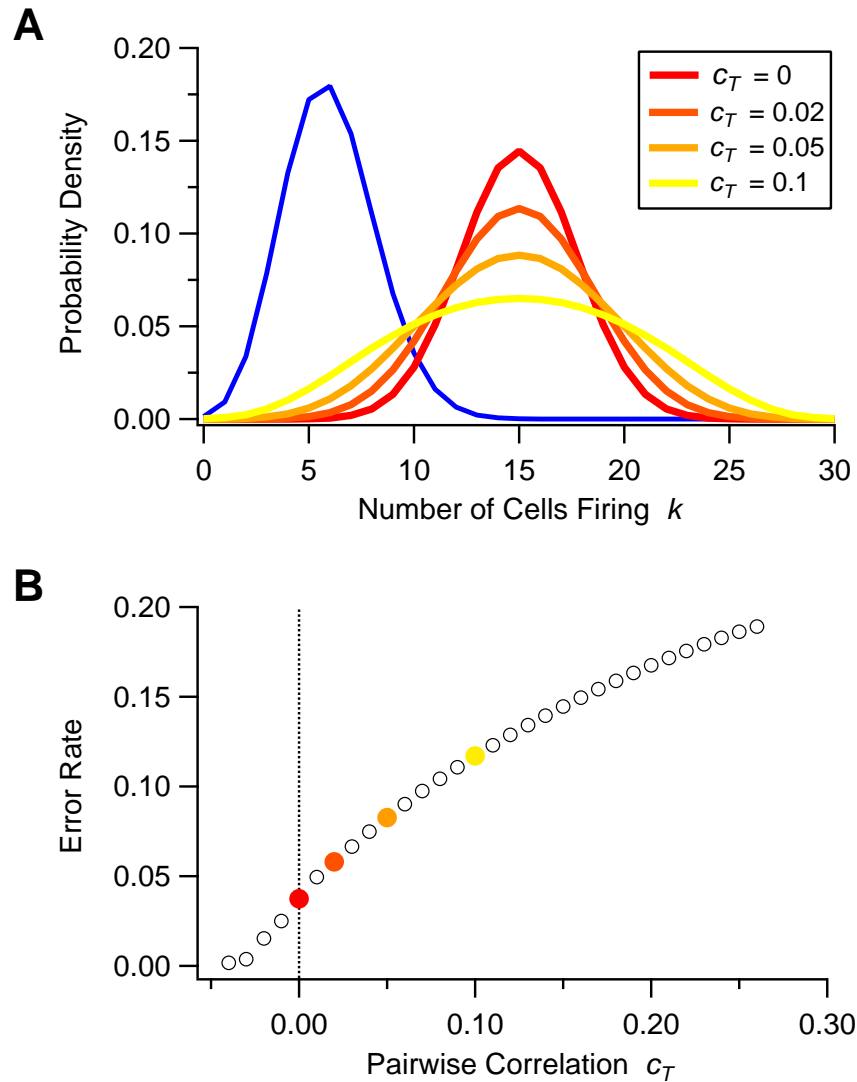}%
\caption{\textbf{(Figure S3.) Homogeneous Populations. A.} Probability
distribution of the spike counts $k$ in a homogeneous population given the
Distracter stimulus (blue) and the Target stimulus with different values of
pairwise correlation, $c_{T}$ (shown by color); parameters are $N=30$ neurons,
$p_{T}=0.5$, $p_{D}=0.2$. \textbf{B.} Probability of error as a function of
the pairwise correlation during the Target stimulus, $c_{T}$ ($N=30$ neurons,
$p_{T}=0.5$, $p_{D}=0.2$, $c_{D}=0$), with examples from panel A (dots with
color matching panel A).}%
\end{center}
\end{figure}

\bigskip

\bigskip
\end{document}